\documentclass[preprint2,times,tighten]{aastex6}
\usepackage{amsmath,amstext}
\usepackage[all]{hypcap} 
\usepackage{multirow}
\usepackage{rotating}

\newcommand{\kms}{km s$^{-1}$}

\newcommand{\sigmol}{\mbox{$\Sigma_{\rm mol}$}}
\newcommand{\sigsfr}{\mbox{$\Sigma_{\rm SFR}$}}
\newcommand{\sigstar}{\mbox{$\Sigma_*$}}
\newcommand{\tdep}{\mbox{$\tau_{\rm dep}$}}

\newcommand{\alphaco}{$\alpha_{\rm CO}$}

\newcommand{\halfmol}{$R_{1/2}^{\rm mol}$}
\newcommand{\halfstar}{$R_{1/2}^*$}
\newcommand{\tauctr}{$\tau_{\rm center}$}
\newcommand{\taudisk}{$\tau_{\rm disk}$}
\newcommand{\taumed}{$\tau_{\rm dep,med}$}

\slugcomment{Accepted for Publication in ApJ}

\received{on April 8, 2017}
\revised{on August 5, 2017}
\accepted{on August 24, 2017}

\shorttitle{Variations in the Molecular Gas Depletion Time in EDGE Galaxies}
\shortauthors{Utomo et al.}

\begin{document}

\title{The EDGE--CALIFA Survey: Variations in the Molecular Gas Depletion Time in Local Galaxies}

\author{Dyas Utomo\altaffilmark{1}, Alberto D. Bolatto\altaffilmark{2}, Tony Wong\altaffilmark{3}, Eve C. Ostriker\altaffilmark{4}, Leo Blitz\altaffilmark{1}, Sebastian F. Sanchez\altaffilmark{5}, Dario Colombo\altaffilmark{6}, Adam K. Leroy\altaffilmark{7},  Yixian Cao\altaffilmark{3}, Helmut Dannerbauer\altaffilmark{8}, Ruben Garcia-Benito\altaffilmark{9}, Bernd Husemann\altaffilmark{10}, Veselina Kalinova\altaffilmark{6}, Rebecca C. Levy\altaffilmark{2}, Damian Mast\altaffilmark{11}, Erik Rosolowsky\altaffilmark{12}, and Stuart N. Vogel\altaffilmark{2}}

\altaffiltext{1}{Department of Astronomy, University of California, Berkeley, CA 94704, USA (email: dyas@berkeley.edu)}
\altaffiltext{2}{Department of Astronomy, University of Maryland, College Park, MD 20642, USA}
\altaffiltext{3}{Department of Astronomy, University of Illinois, Urbana, IL 61801, USA}
\altaffiltext{4}{Department of Astrophysical Sciences, Princeton University, Princeton, NJ 08544, USA}
\altaffiltext{5}{Instituto de Astronom\'\i a, Universidad Nacional Aut\'onoma de M\'exico, A.P. 70-264, 04510 M\'exico, D.F.,  Mexico}
\altaffiltext{6}{Max Planck Institute for Radio Astronomy, Auf dem H\"ugel 69, D-53121 Bonn, Germany}
\altaffiltext{7}{Department of Astronomy, The Ohio State University, Columbus, OH 43210, USA}
\altaffiltext{8}{Instituto de Astrof\'{\i}sica de Canarias, E-38205 La Laguna, Tenerife, Spain}
\altaffiltext{9}{Instituto de Astrof\'{\i}sica de Andaluc\'{\i}a, CSIC, E-18008 Granada, Spain}
\altaffiltext{10}{European Southern Observatory, D-85748 Garching bei M\"unchen, Germany}
\altaffiltext{11}{Observatorio Astron\'omico de C\'ordoba, 5000 C\'ordoba, C\'ordoba, Argentina}
\altaffiltext{12}{Department of Physics, University of Alberta, Edmonton, T6G 2E1, Canada}

\begin{abstract}
We present results from the EDGE survey, a spatially resolved CO(1--0) follow-up to CALIFA, an optical Integral Field Unit (IFU) survey of local galaxies. By combining the data products of EDGE and CALIFA, we study the variation in molecular gas depletion time (\tdep) on kiloparsec scales in 52 galaxies. We divide each galaxy into two parts: the center, defined as the region within $0.1 \ R_{25}$, and the disk, defined as the region between $0.1$ and $0.7 \ R_{25}$. We find that 14 galaxies show a shorter \tdep\ ($\sim 1$ Gyr) in the  center relative to that in the disk (\tdep\ $\sim 2.4$ Gyrs), which means the central region in those galaxies is more efficient at forming stars per unit molecular gas mass. This finding implies that the centers with shorter \tdep\ resemble the intermediate regime between galactic disks and starburst galaxies. Furthermore, the central drop in \tdep\ is correlated with a central increase in the stellar surface density, suggesting that a shorter \tdep\ is associated with molecular gas compression by the stellar gravitational potential. We argue that varying the CO-to-H$_2$ conversion factor only exaggerates the central drop of \tdep.
\end{abstract}

\maketitle

\section{Introduction}

Galactic stellar masses grow through a combination of mergers and the formation of stars from their gas reservoir over cosmic time. Therefore, the star formation rate (SFR) is an important factor in driving galaxy evolution \citep[e.g.,][]{kennicutt98b,mckee07,kennicutt12}. In general, star formation involves two processes: (1) the conversion of diffuse, atomic gas into molecular gas in well-shielded regions of high density, and (2) the dynamical collapse of self-gravitating regions within the molecular component to form stars. In galactic regions with low mean gas volume and low surface density, local gas compression by spiral arms or self-gravity may be needed for molecules to form, whereas in galactic regions of high mean gas volume and surface density, most of the gas already molecular \citep[e.g., in M51;][]{schinnerer13}. In this paper, we focus on the second part of the star formation processes, specifically, we study how the relation between molecular gas and SFR changes between the galactic centers and the disks.

In a simple-minded picture, stars form from the gas that contracts under its self-gravity. Naively, one would expect that the relevant time-scale of this process is the free-fall time ($\tau_{\rm ff}$) of the total gas (atomic and molecular), which is inversely proportional to the square-root of gas volume density ($\rho_{\rm gas}^{-0.5}$). The implication of this simple scenario is that SFR relates to the amount of gas as $\rho_{\rm SFR} \propto \rho_{\rm gas}/\tau_{\rm ff} \propto \rho_{\rm gas}^{1.5}$.\footnote{The other time scales that are often used in literature are the orbital time $\Omega^{-1}$ \citep[e.g.,][]{elmegreen97,silk97}, where $\Omega$ is the angular speed of the disk, and the vertical time $H/\sigma$ \citep{ostriker10,ostriker11}, where $H$ and $\sigma$ are the thickness and velocity dispersion of the gas.} In general, the relation between SFR and total gas density is called the Kennicutt-Schmidt (KS) relation, after the seminal papers by \citet{schmidt59} and \citet{kennicutt98}.\footnote{Actually, \citet{schmidt59} proposed $\rho_{\rm SFR} \propto \rho_{\rm gas}^2$ and \citet{kennicutt98} found $\Sigma_{\rm SFR} \propto \Sigma_{\rm gas}^{1.4}$, where $\Sigma$ is the surface density. Since $\Sigma$ is the integration of $\rho$ along the projected disk thickness, the translation from $\rho_{\rm SFR}$ to $\Sigma_{\rm SFR}$ depends on the scale height of the ISM.}

Observations in the local universe show that stars form in molecular clouds, so we expect that SFR correlates better with the amount of molecular gas, rather than the total amount of atomic plus molecular gas \citep[e.g.,][]{wong02,kennicutt07,bigiel08}. Even though the molecular phase may itself not be necessary to form stars \citep{glover12}, molecular gas that forms under the high-density conditions are also favorable to gravitational collapse, thus giving rise to a strong KS relation \citep{krumholz11}. For simplicity, in this paper we refer to the relationship between SFR and molecular gas surface densities as the KS relation.

Resolved studies of nearby galaxies found that the correlation between SFR and molecular gas surface densities is approximately linear\footnote{There is a tension on the actual slope of KS relation. For example, \citet{kennicutt07} derived a slope of 1.37 in M51, while \citet{bigiel08} derived a slope of 0.84 in the same galaxy. There are two possible reasons of this difference. (1) Different treatments on the background radiation that is used as a tracer for SFR \citep{liu11,calzetti12}. A removal of background radiation leads to a steeper slope. (2) Different regions in M51 have different slopes of KS relation \citep{leroy17}, so that the derived slope depends on which regions have larger weight in the best-fit slope.} in galaxy disks, with $\Sigma_{\rm SFR} \propto \Sigma_{\rm mol}$ on  kiloparsec (kpc) scales for surface densities $\Sigma_{\rm H2} \gtrsim 3 \ M_{\odot} \ {\rm pc}^{-2}$ over a wide range of local environments \citep[e.g.,][]{bigiel08,leroy08}. Furthermore, in nearby galaxies, the near-linear molecular KS relation extends to the low metallicity regime \citep[$Z/Z_{\odot} \approx 0.2$;][]{bolatto11,jameson16} and to the outer part of galaxies, where the gas surface density is low and atomic dominated \citep{schruba11}. A possible reason for this widespread relationship is that the properties of molecular clouds are similar from one galaxy and region to another \citep{bolatto08}, so that GMCs convert the molecular gas into stars at the same rate.

For most of the gas in normal galaxies, the linearity of KS relation implies the molecular gas depletion time, defined as $\tau_{\rm dep} \equiv \Sigma_{\rm mol}/\Sigma_{\rm SFR}$, is approximately constant, with a typical value of $2.2$ Gyrs in nearby galaxies \citep[e.g.,][]{bigiel08,leroy08,rahman12,leroy13}. Loosely, we interpret $\tau_{\rm dep}$ as the time scale to convert all molecular gas reservoir in a galaxy (or a given region within a galaxy) into stars at the current SFR. The fact that $\tau_{\rm dep}$ is less than the Hubble time implies that galaxies need to replenish their molecular gas reservoir through stellar feedback (e.g., supernovae, stellar winds, AGB stars, and planetary nebula), conversion from atomic to molecular gas, and accretion from the intergalactic medium or from satellite galaxies \citep[e.g.,][]{genzel10,bauermeister10,lilly13}. However, direct observational signature of this accretion is still challenging.

Despite the current evidence towards the linearity of KS relation, there are, at least, three regimes where this linearity breaks down: (1) in the ULIRGs and starburst galaxies, i.e. galaxies above the star forming main-sequence \citep[e.g.,][]{daddi10,genzel10,genzel15}, (2) at resolution finer than $\sim 500$ pc \citep[e.g.,][]{schruba10,calzetti12,kruijssen14}, and (3) in galactic centers \citep[e.g.,][]{jogee05,leroy13}. In addition, a trend of \tdep\ with respect to stellar mass on galaxy-by-galaxy basis was reported by \citet{saintonge11b} in COLDGASS sample and \citet{bolatto17} in EDGE sample.

The steeper-than-linear molecular KS relation in regions of very high molecular surface density has been interpreted as a result of higher molecular gas pressure \citep{ostriker11} and density \citep{krumholz12}.  Higher pressure requires a higher star formation rate per unit molecular mass to offset enhanced turbulent dissipation and cooling, and higher density is associated with shorter dynamical times, which control gravitational contraction.

This paper is based on the combination of the CO data from the EDGE survey \citep{bolatto17} and the optical IFU data from the CALIFA survey \citep{sanchez12}. In the first EDGE paper by \citet{bolatto17}, we showed that the relation between \sigsfr\ and \sigmol\ is approximately linear, with a separation of \tdep\ between high and low masses galaxies. We extend that study in this paper by analyzing the variations of \tdep\ between galactic centers and disks, with a goal to quantify and understand the cause of those variations and their implications in galaxy evolution.

This paper is organized as follows. Overviews of the EDGE and CALIFA data products and the sample selection are described in $\S$\ref{sec:2} and $\S$\ref{sec:3}, respectively. Then, in $\S$\ref{sec:4} we compare $\tau_{\rm dep}$ in the centers relative to those in the disk. Specifically, we investigate whether the difference of \tdep\ between the centers and the disks is due to SFR, molecular gas, or stellar surface density. In $\S$\ref{sec:5}, we discuss the effect of the CO-to-H$_2$ conversion factor, the connection between \tdep\ and oxygen abundance, the size of molecular and stellar disks, and the possibility that the galactic center undergoes cycles of star formation. Lastly, we summarize our findings in $\S$\ref{sec:6}. All logarithms in this paper are base 10 logarithms.

\section{Data Descriptions} \label{sec:2}

\subsection{The EDGE Survey}

\begin{figure*}
\begin{center}
\includegraphics[width=1\textwidth]{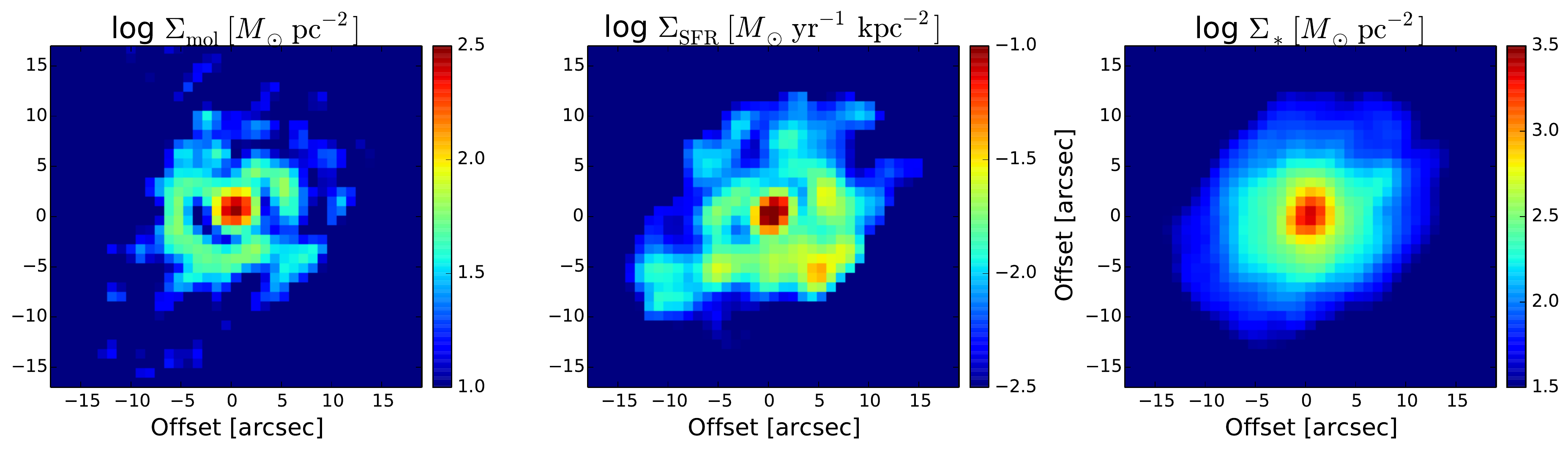}
\end{center}
\caption{An example of the EDGE-CALIFA data products. From left to right: the maps of molecular gas, SFR, and stellar mass surface density of NGC 2253.}
\label{fig:ngc2253}
\end{figure*}

The EDGE survey targets 126 galaxies in the CO(1--0) and $^{13}$CO(1--0) lines using the CARMA observatory \citep{bock06} in the D and E arrays from 2014 October until 2015 May. The observational details and data reductions of the EDGE survey are presented in \citet{bolatto17}. Briefly, the EDGE samples are selected from the CALIFA Second Data Release \citep{garcia-benito15} based on their fluxes in WISE 22$\mu$m band \citep{wright10}. The raw data are reduced using the {\tt MIRIAD} package \citep{sault95} into data cubes (i.e. CO intensity in velocity and two-dimensional spaces) using an automated pipeline based on scripts developed for the STING galaxy survey \citep{rahman12,wong13}.

The beam size of each galaxy varies with a typical value of $4\arcsec.5$, which corresponds to a median physical scale of about 1.5 kpc. This physical resolution is slightly larger than previous CO surveys, such as BIMA SONG \citep[$\sim 360$ pc;][]{helfer03}, HERACLES \citep[$\sim 500$ pc;][]{leroy09}, and STING \citep[$160 - 1250$ pc;][]{rahman12}, because our sample covers farther median distance than those surveys. The pixel size is $2'' \times 2''$. The velocity resolution is 10 \kms\ with a typical velocity range of 860 \kms, thus, it covers out to the flat part of the rotation curve where CO is detected. The data cubes that provide an estimate of $1\sigma_{\rm rms}$ noise level at each pixel were also generated during the data reduction processes.

In order to separate signal from noise, we create masks through the following steps in {\tt IDL} \citep[code available at \url{https://github.com/tonywong94/idl_mommaps};][]{wong13}. First, we smooth the data into $9\arcsec$ resolutions with a Gaussian kernel. The aim of this smoothing is to reach a higher signal to noise ratio (SNR). Then, we search for contiguous regions, starting from pixel that has SNR $\geq 3.5$ down to regions that have SNR $= 2$. The aim of contiguous regions is to remove noise that has high SNR by chance, but only localized into one to few pixels \citep[e.g.,][]{rosolowsky06}. An additional padding of 2 pixels surrounding the $2\times$SNR contours are added into the mask to capture low level emission. Finally, we apply these masks to the data cubes in their original resolutions ($4\arcsec.5$ and 10 km s$^{-1}$). We define these contiguous regions, including the padding, as masked regions.

The masked data cubes are integrated along the velocity axis to get the CO surface brightness maps (zeroth moment maps). Similarly, the uncertainties of the maps are taken by integrating the estimated noise along the velocity axis within the masked cubes. In the analyses, we use these uncertainty maps as $1\sigma_{\rm rms}$ noise level. Note that not all masked CO surface brightness maps are higher than $2\sigma_{\rm rms}$ level, therefore, we treat emissions below $2\sigma_{\rm rms}$ level as non-detections, even though these emissions are located within the mask.

We convert the CO surface brightness and its uncertainty maps into molecular gas surface density ($\Sigma_{\rm mol}$) maps by using a constant CO-to-H$_2$ conversion factor ($\alpha_{\rm CO}$) of 4.4 $M_{\odot} \ {\rm pc}^{-2}$ (K km s$^{-1}$ pc$^2$)$^{-1}$, including the mass contribution from Helium. In general, $\alpha_{\rm CO}$ can vary as a function of metallicities and stellar surface densities \citep{bolatto13}. In our approach, we take a Galactic value of $\alpha_{\rm CO}$, and then, we consider how the variations of $\alpha_{\rm CO}$ affect our results in $\S$\ref{sec:alpha_co}. Note that any surface densities measurement has been corrected (deprojected) from inclination $(i)$ by using a correction factor of cos$(i)$. An example of the map of \sigmol\ is shown as the left panel of Figure~\ref{fig:ngc2253}.

\subsection{The CALIFA Survey}

CALIFA is an optical Integral Field Unit (IFU) survey of $\sim 600$ local galaxies at the redshift range of $0.005 \lesssim z \lesssim 0.03$ using the 3.5-m telescope at the Calar-Alto observatory \citep{sanchez12}. The CALIFA samples are selected from the SDSS DR7 database \citep{abazajian09} based on their diameter in $r-$band ($45'' < D_{25} < 80''$), so that they fit well within the IFU field-of-view of $1'.3$, or equivalently $\sim 2.5$ effective radius \citep{walcher14}, but statistically still represents the population of $z \sim 0$ galaxies in the color-magnitude diagram. In an IFU survey, we can get spatial and spectral information of an object, simultaneously. The spatial resolution of CALIFA is $\sim 2''.5$ (or $\sim$ kpc scale) and the spectral range of CALIFA covers 3700 to 7000 \AA, so that it captures the stellar absorption lines and the nebular emission lines.

We take the following additional steps to create homogeneous datasets between EDGE and CALIFA. (1) Recenter any offset in CALIFA data by using cross-correlation between CALIFA $V$-band and SDSS $g$-band images. In general, the offsets are about few arcsec and not systematic. (2) Regrid the CALIFA data by using {\tt MIRIAD} task {\tt regrid}, so that it has the same spatial coordinate as in the EDGE data with a common pixel size of $2\arcsec \times 2\arcsec$. In this process, we also degrade the resolution of CALIFA images to match the resolution of EDGE images by using {\tt MIRIAD} task {\tt convol}. The total flux is conserved during those processes. (3) Blanking the CALIFA data that are contaminated by foreground stars and neighboring galaxies. (4) Separating signals from noise by blanking any pixels that have SNR $< 2$, where we use the median-absolute-deviation of the CALIFA image as an estimate of the noise. As in the EDGE dataset, all surface densities derived from the CALIFA dataset have been corrected by cos$(i)$ to take into account the size deprojection due to inclination.

\subsubsection{The Star Formation Rate Surface Density}

\begin{figure*}
\begin{center}
\includegraphics[width=0.85\textwidth]{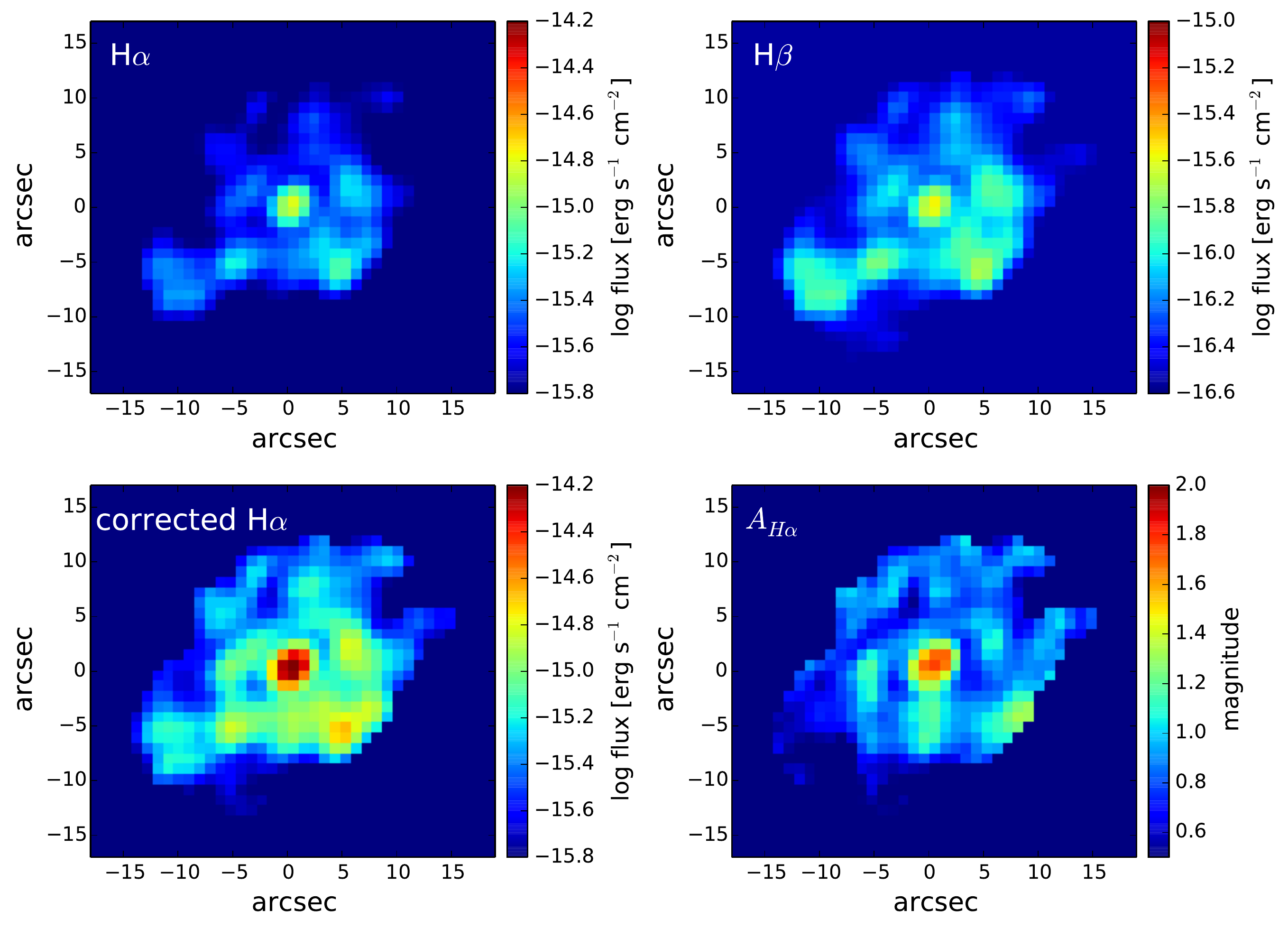}
\end{center}
\caption{An application of Balmer decrement method to NGC 2253. Top left: H$\alpha$ fluxes. Top right: H$\beta$ fluxes. Bottom right: Dust extinction at H$\alpha$ ($A_{H\alpha}$). Bottom left: By correcting $A_{H\alpha}$ to H$\alpha$ fluxes reveals the dust corrected H$\alpha$ fluxes.}
\label{fig:balmer_dec}
\end{figure*}

The post-processing results of CALIFA data \citep[{\tt Pipe3D} version 2.2 from][]{sanchez16} provide the intensity maps of emission lines, such as H$\alpha$ and H$\beta$. To derive maps of the SFR surface density ($\Sigma_{\rm SFR}$), first, we calculate the nebular extinction at H$\alpha$ wavelength, $A_{{\rm H}\alpha}$, by utilizing the ratio of H$\alpha$ and H$\beta$ fluxes \citep[Balmer decrement method; e.g.,][]{dominguez13} and compare it with its intrinsic value (zero extinction) of 2.86 \citep[for case B recombination at temperature of $10^4$ K and electron density of 100 cm$^{-3}$;][]{osterbrock89}. In the process, we also use a Galactic extinction curve \citep{cardelli89} with $R_V = 3.1$. The result will be similar if we use \citet{calzetti00} extinction curve with $R_V = 4.1$, because $A_{{\rm H}\alpha,{\rm Calzetti}}/A_{{\rm H}\alpha,{\rm Cardelli}} = 1.03$ \citep{catalan15}. The resulting pixel-by-pixel mean value of $A_{{\rm H}\alpha}$ is about 1 magnitude. Then, we apply this $A_{{\rm H}\alpha}$ to H$\alpha$ maps to get the {\it dust-corrected} (or extinction-free) H$\alpha$ maps. An example of this Balmer decrement method is shown in Figure~\ref{fig:balmer_dec}.

We convert the dust-corrected H$\alpha$ maps to the SFR surface density maps following the prescriptions in \citet{calzetti07}, based on a stellar population model with 100 Myr of constant SFR, solar metallicity, and an IMF that has a slope of $-1.3$ within $0.1 < M_*/M_{\odot} < 0.5$ and a slope of $-2.3$ within $0.5 < M_*/M_{\odot} < 120$ stellar mass range. The IMF for this SFR prescription is similar to a \citet{kroupa01} IMF, which is a factor of 1.59 smaller than those derived from a \citet{salpeter55} IMF within mass range of $0.1 - 100 \ M_\odot$ \citep{madau14}. An example of the \sigsfr\ maps is shown as the second column of Figure~\ref{fig:ngc2253}.

\begin{figure}
\begin{center}
\includegraphics[width=0.48\textwidth]{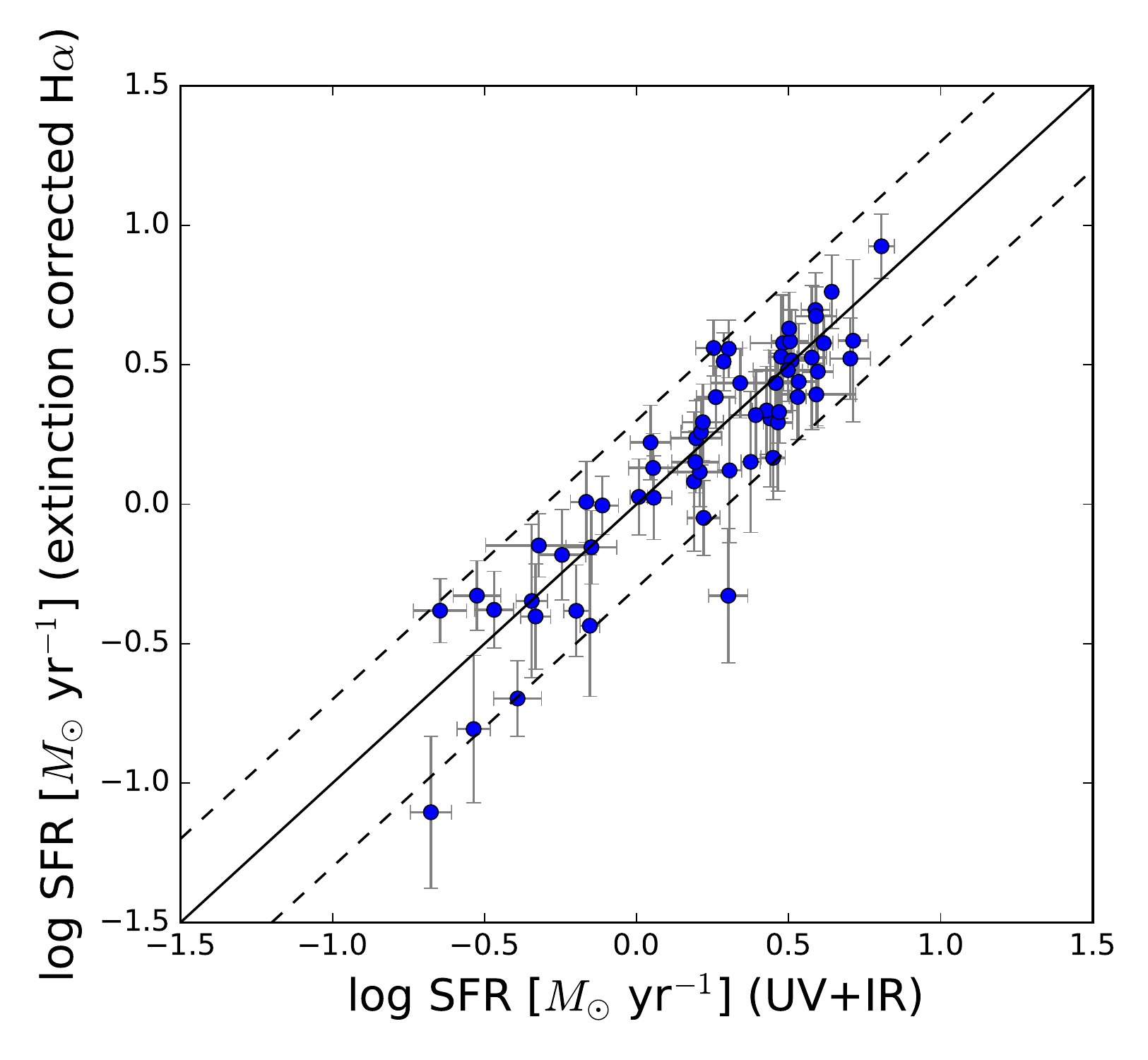}
\end{center}
\caption{A comparison of SFR measurements from extinction-corrected H$\alpha$ (this paper) and UV+IR from \citet{catalan15}. We apply aperture correction for our SFR measurement as suggested by \citet{catalan15}. The solid line is the one-to-one relationship, while the dashed line is 0.3 dex away from the solid line. The uncertainties of SFR measurement in this paper is calculated using the error propagation from the uncertainties in H$\alpha$ and H$\beta$ measurements. A 20\% uncertainty due to SFR calibration \citep{calzetti07} has been included as well.}
\label{fig:test_sfr}
\end{figure}

As a check, we compare the SFR of {\it extinction-corrected} H$\alpha$ emission that we derived above with the SFR derived from the ultraviolet (UV) emission plus total-infrared (TIR) emission from \citet{catalan15}. The UV emission traces the unobscured SFR, while the TIR emission compensates for the obscured SFR that is reradiated by dust. We do galaxy-by-galaxy comparisons by integrating our resolved SFR because the infrared data are unresolved. Since the H$\alpha$ emission is more extended than the FoV of CALIFA survey, we apply an aperture correction of 1.4 as suggested by \citet{catalan15}. In Figure~\ref{fig:test_sfr}, we show that both measurements are in agreement within a factor of $\sim 2$.

\subsubsection{The Gas-phase Metallicities} \label{sec:metals}

We determine the gas-phase metallicities by using emission lines ratio of O\textsc{iii}[5007\AA]/H$\beta$ and N\textsc{ii}[6583\AA]/H$\alpha$ \citep[i.e. the O3N2 method;][]{alloin79,pettini04}. We use the following prescription from \citet{marino13}
\begin{equation}
12 + {\rm log(O/H)} = 8.533 - 0.214 \ {\rm log} \left( \frac{\rm O\textsc{iii}}{{\rm H}\beta} \frac{{\rm H}\alpha}{\rm N\textsc{ii}} \right).
\end{equation}
The coefficient of this method has been calibrated by using the electron temperature based measurements in 603 H\textsc{ii} regions extracted from literatures and 3423 additional H\textsc{ii} complexes from the CALIFA survey. The resolved metallicities in our sample range from 8.3 to 8.6, slightly below the Solar metallicity of 8.7 \citep{prieto01}.

\subsubsection{The Stellar Ages and Mass Surface Densities}

We take the luminosity-weighted, stellar population ages and the dust-corrected, stellar mass surface densities (\sigstar) from the data products of {\tt Pipe3D} version 2.2 \citep{sanchez16}. Briefly, the data products are derived from the best fit of stellar spectra from a combination of the GRANADA \citep{martins05} and MILES libraries \citep{sanchez-blazquez06,vazdekis10,falcon-barroso11}, that cover 39 grids of stellar ages (from 1 Myr to 13 Gyrs) and 4 grids of stellar metallicities ($Z/Z_{\odot} = 0.2, 0.4, 1$ and 1.5). We convert the \sigstar\ maps from a \citet{salpeter55} IMF to a \citet{kroupa01} IMF by dividing it by a factor of 1.59 \citep{madau14}.

\section{Sample Selection} \label{sec:3}

We select 52 galaxies from 126 EDGE galaxies based on the following three criteria. (1) They are not dominated by AGN and LINER. (2) They have sufficient SFR and CO detection that cover both the centers and the disk. (3) The inclination $(i)$ is less than $75^{\circ}$. The inclinations are taken from the following sources, ordered by priority: (1) the best fit of CO rotation curve, whenever it is possible (Levy et al. in preparation), (2) from the shape of the outer isophote, or (3) from the HyperLEDA catalog \citep{makarov14}. A list of the galaxy sample is tabulated in Appendix~\ref{app:gal_props}.

We exclude AGN and LINER emission regions based on N\textsc{ii}/H$\alpha$ and O\textsc{iii}/H$\beta$ line ratios \citep[i.e. the BPT diagram;][]{baldwin81,kewley02,kauffmann03}. Any data points above the demarcation line of \citet{kewley02} are blanked. We also blank any regions that have H$\alpha$ equivalent width less than 6 \AA, because $\sim 80\%$ of stars in those regions are older than $\sim 500$ Myrs, and hence, not associated to star forming regions \citep{sanchez14}. Note that the LINER emission region are not only concentrated in the center, but also in the disk, possibly due to photo-ionization from AGB stars \citep{singh13,belfiore16}. A galaxy is removed from the samples if all pixels in the center (i.e. within $0.1 \ R_{25}$) is AGN/LINER-like emission. Based on that criterion, 31 galaxies from the EDGE sample are removed. 

We further remove 17 galaxies that do not have sufficient CO or SFR detection in the centers or in the disks, because measurement of \tdep\ is severely contaminated by non-detection. If a galaxy has less than 2 detected pixels in the center or in the disk, then that galaxy is removed from the sample. Lastly, 26 galaxies with $i \gtrsim 75^\circ$ (equivalents to the ratio of minor to major axis of less than 0.25) are removed because highly inclined galaxies yield few sampling points along the minor axis, resulting in a deprojected beam elongated parallel to the minor axis in the plane of the galaxy, and high uncertainty in the estimation of dust extinction.

Our final sample has stellar masses ($M_*$) from $4 \times 10^9$ to $2 \times 10^{11} \ M_\odot$, molecular gas masses ($M_{\rm mol}$) from $8 \times 10^7$ to $1 \times 10^{10} \ M_\odot$, and gas-phase metallicities ($12+$log[O/H]) from 8.4 to 8.6 dex. Our sample consists of 50 spirals (Hubble type from Sa to Sd) and 2 early-types, which 24 of them are barred and 7 of them are interacting \citep{barrera-ballesteros15}. The ranges in the stellar and molecular gas masses are comparable to the unresolved survey of COLDGASS \citep{saintonge11a,saintonge11b}. In addition, we have a comparable number of galaxies and cover farther distance in the local volume ($26 \lesssim d \lesssim 169$ Mpc) than previous resolved surveys, such as BIMA SONG \citep[44 galaxies; $2 \lesssim d \lesssim 26$ Mpc;][]{helfer03}, Nobeyama CO Survey \citep[40 galaxies; $d < 25$ Mpc;][]{kuno07}, CARMA STING \citep[14 galaxies; $5 \lesssim d \lesssim 43$ Mpc;][]{rahman12}, JCMT NGLS \citep[155 galaxies; $d < 25$ Mpc;][]{wilson12}, and HERACLES \citep[48 galaxies; $3 \lesssim d \lesssim 15$ Mpc;][]{leroy13,schruba12}. Thus, our sample bridges the gap between nearby and higher redshift galaxies.

\section{Results} \label{sec:4}

\begin{figure}
\begin{center}
\includegraphics[width=0.48\textwidth]{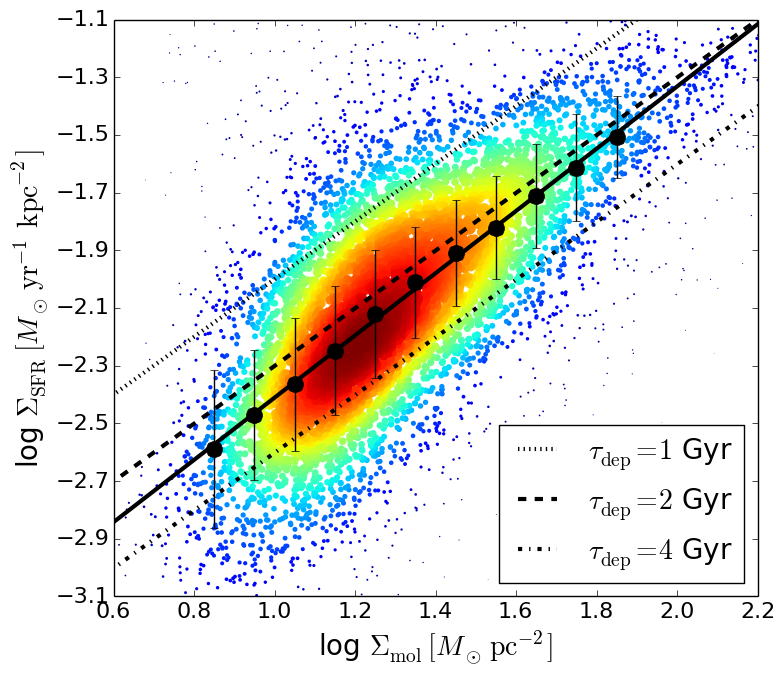}
\end{center}
\caption{The relationship between \sigmol\ and \sigsfr\ for 52 galaxies in our sample. The data points are pixel-by-pixel measurements ($2'' \times 2''$), with colors and point sizes are coded by the density of data points. The black dots are the median value of \sigsfr\ within bins of \sigmol. A linear fit to the black dots is given by the solid black line. This linear fit has a slope of $1.08 \pm 0.01$ and an intercept point of $-3.49 \pm 0.02$. The dotted, dashed, and dash-dotted lines correspond to \tdep\ = 1, 2, and 4 Gyrs, respectively.}
\label{fig:kslaw}
\end{figure}

In Figure~\ref{fig:kslaw}, we show the KS relation for molecular gas. The data points are from pixel measurements (detected both in SFR and CO) in 52 galaxies. The median values of \sigsfr\ for a given bin of \sigmol\ are marked as black dots, while the constant values of \tdep\ = 1, 2, and 4 Gyrs are indicated. There is a tendency that the high \sigmol\ region (top right in Figure~\ref{fig:kslaw}) has a slightly shorter \tdep\ than the low \sigmol\ region (i.e. the best-fit slope is slightly larger than unity). Since galactic centers have higher \sigmol\ than that in the disks, this indicates that the centers have shorter \tdep\ than in the disks.

In order to study the variation of $\tau_{\rm dep}$ between the galactic centers and disks, we need to separate the central region of a galaxy. To do so, we define the center as a region within $0.1 \ R_{25}$ from the galactic nucleus, and the disk as a region between $0.1 \ R_{25}$ and $0.7 \ R_{25}$. Therefore, \tauctr\ and \taudisk\ are the median of \tdep\ over all detected pixels in the center and in the disk, respectively. If the median or the whole value of \tdep\ in a galaxy is used, it means we cover both the center and the disk, and we refer to it as \taumed. If the number of detected pixels in the disks is much larger than those in the centers, then the values of \taumed\ is similar to \taudisk. We adopt $0.7 \ R_{25}$ as the outermost radius because CO is hardly detected beyond that radius.

The radial distance to the galactic nucleus is calculated using the assumption that the molecular gas lies on the galactic mid-plane, without warp, isophotal twist, and misalignment. Since each galaxy has different physical size in kpc, sometimes we normalize the radius with respect to $R_{25}$, i.e. the radius where the surface brightness is 25 mag arcsec$^{-2}$ in the $B-$band. We adopt the values of $R_{25}$ from the HyperLEDA catalog. The scaling relation between $R_{25}$ and the stellar scale length ($l_*$) is $R_{25} = (4.6 \pm 0.8) \ l_*$ \citep{leroy08}. Unless otherwise stated, throughout this paper we focus on the star forming regions detected in both CO (\sigmol\ $\gtrsim 10 \ M_{\odot}$ pc$^{-2}$) and H$\alpha$ in pixel-by-pixel basis ($\sim$ kpc scale).

\subsection{Depletion Time in the Centers and in the Disks}

\begin{figure}
\begin{center}
\includegraphics[width=0.48\textwidth]{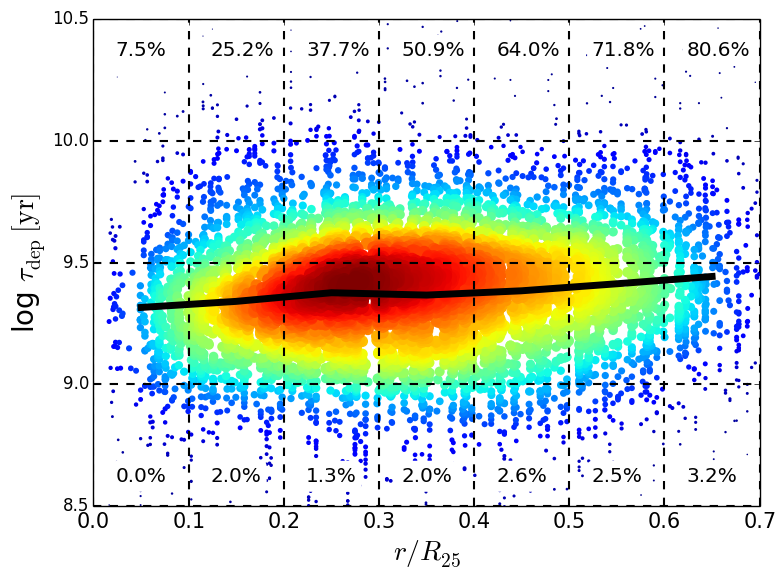}
\end{center}
\caption{The depletion time as a function of radius, aggregated over all detected regions in the sample. The data points are the pixel-by-pixel ($2\arcsec \times 2\arcsec$) measurements. The colors and sizes of points represent the global density of the data points and the solid line is the median value of \tdep\ in radial bins. On the top and bottom of the figure, we label the fractions of non-detection pixels that correspond to upper and lower limits in \tdep, respectively. Upper limits in \tdep\ are pixels with known SFR but CO is not detected, and vice versa for lower limits. The H$\alpha$ measurements are more sensitive than the CO maps, therefore, the fractions of upper limits are higher than the fractions of lower limits at any radius.}
\label{fig:tdep_rad}
\end{figure}

Since CO emission is patchy, not all regions within a galaxy are detected in CO and H$\alpha$. To accrue more signal-to-noise and get a better radial coverage across the sample, we aggregate the \tdep\ measurements as a function of $r/R_{25}$ for all galaxies. By doing this measurement for the CO detections only we focus on regions that, like most galaxy centers, are dominated by molecular gas ($\sigmol\ge10$~M$_\odot$\,pc$^{-2}$), and where similar star-formation mechanisms are likely to operate. In Figure~\ref{fig:tdep_rad}, \tdep\ in each detected pixels are plotted as a function of radius. The median value of \tdep\ is 2.4 Gyrs with $\sim 0.5$ dex scatter. This value is in line with the previous measurements in nearby galaxies \citep[e.g.,][]{rahman12,bigiel11,leroy13}. Pointings in the center, however, have shorter \tdep\ than those in the disk. However, the dip of \tauctr\ does not occur in all galaxies in the sample, and becomes more prominent when we separate those galaxies from the rest of the sample (see $\S$~\ref{sec:classify}).

\begin{figure*}[!t]
\begin{center}
\includegraphics[width=1\textwidth]{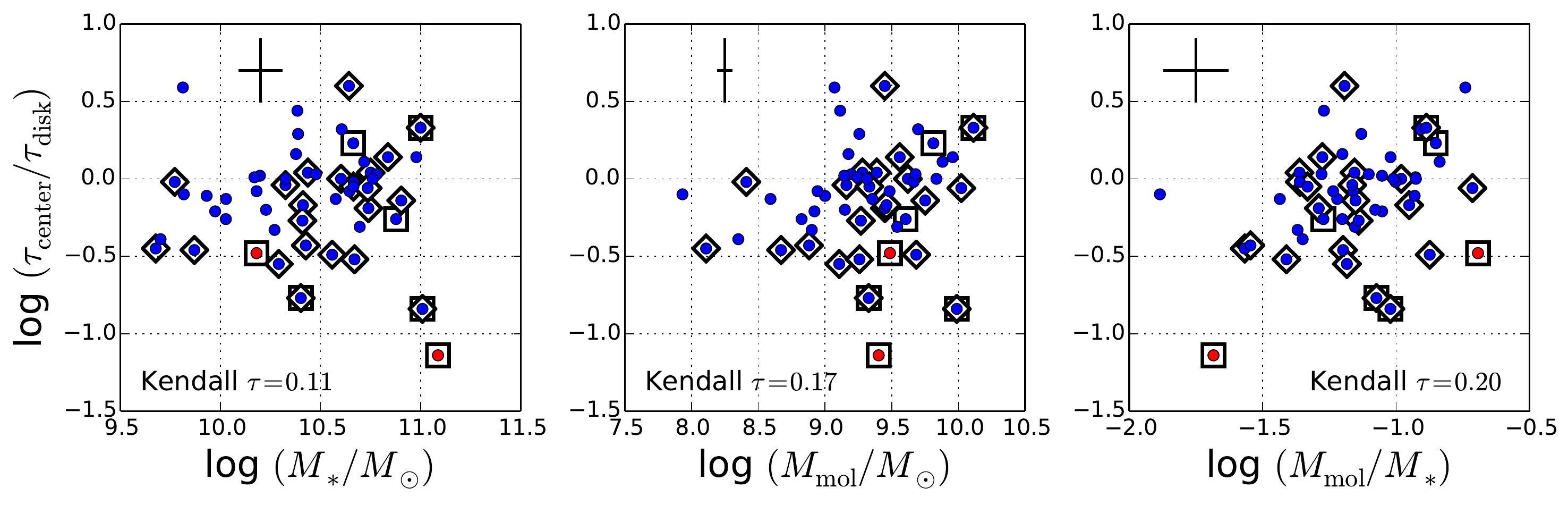}
\end{center}
\caption{The values of \tdep\ in the centers, relative to those in the disks, are plotted against the global parameters of galaxies: stellar masses (left panel), molecular gas masses (middle panel), molecular-to-stellar mass ratio (right panel). Blue and red points mark the late-type and early-type galaxies, respectively. The diamonds mark the barred galaxies, while the squares mark the interacting galaxies. The values of \citet{kendall38} $\tau-$coefficient that measure the rank correlation between two quantities are indicated in each panel. A perfect correlation has a $\tau-$coefficient of 1, while two independent quantities have a $\tau-$coefficient of 0. Since the correlations are not significant, we do not plot the best-fit line. The crosses represent the typical error bars of the data points.}
\label{fig:tdep_global}
\end{figure*}

In Figure~\ref{fig:tdep_global}, \tauctr\ and \taudisk\ for each galaxy are shown. The ratio between \tauctr\ and \taudisk\ in our sample can reach a factor of $\sim 10$, but the ratio in most galaxies is between unity and a factor of 3. The scatter in log(\tauctr$/$\taudisk) is larger in the high stellar and molecular gas masses regime. We investigate whether the variation of \tauctr\ relative to \taudisk\ is correlated to the global properties of galaxies, namely the stellar masses $(M_*)$, the molecular gas masses $(M_{\rm mol})$, the Hubble types, the gas-phase metallicities, and the age of stellar populations. We adopt RC3 \citet{devaucouleurs91} indices from the HyperLEDA catalog as morphological types. For the oxygen abundance and the age of stellar population, we use their median value within $1 \pm 0.2$ effective radius ($R_e$) because \citet{sanchez16} suggest that the value at $R_e$ is a good representation for a galaxy.

We do not find correlation between log($\tau_{\rm center}/\tau_{\rm disk}$) and morphology, gas-phase metallicity, or age of stellar populations at $R_e$, probably because we have limited range in morphology (96\% of our samples are spirals) and gas phase metallicity (only $\sim 0.2$ dex of variations). Furthermore, the age of stellar populations at $R_e$ reflect the value in the disks, where \taudisk\ does not vary as much as \tauctr. If we measure the stellar age in the center, however, galaxies with low values of log($\tau_{\rm center}/\tau_{\rm disk}$) have younger ages for stellar populations (see $\S$\ref{sec:burst}). There is also no significant correlation between \tauctr$/$\taudisk\ and $M_*$, $M_{\rm mol}$, and $M_{\rm mol}/M_*$ (Figure~\ref{fig:tdep_global}), as indicated by low values of \citet{kendall38} $\tau-$coefficient.

It should be noted that three galaxies with the lowest values of log(\tauctr$/$\taudisk) are interacting galaxies (marked as black squares in Figure~\ref{fig:tdep_global}). In addition, barred galaxies, marked as black diamonds in Figure~\ref{fig:tdep_global} \citep[identified from the photometric fit of][or from the HyperLEDA catalog]{mendez-abreu17}, tend to have lower values of log(\tauctr$/$\taudisk) than unbarred galaxies. The mean values of log(\tauctr$/$\taudisk) for interacting and barred galaxies are $-0.42 \pm 0.51$ and $-0.22 \pm 0.28$, while the corresponding value for unbarred galaxies is $-0.03 \pm 0.35$. This indicates that perturbed systems may enhance the star formation efficiency in the center.

\subsection{Separations of Galaxies into Three Groups of \tdep} \label{sec:classify}

\begin{figure*}
\begin{center}
\includegraphics[width=\textwidth]{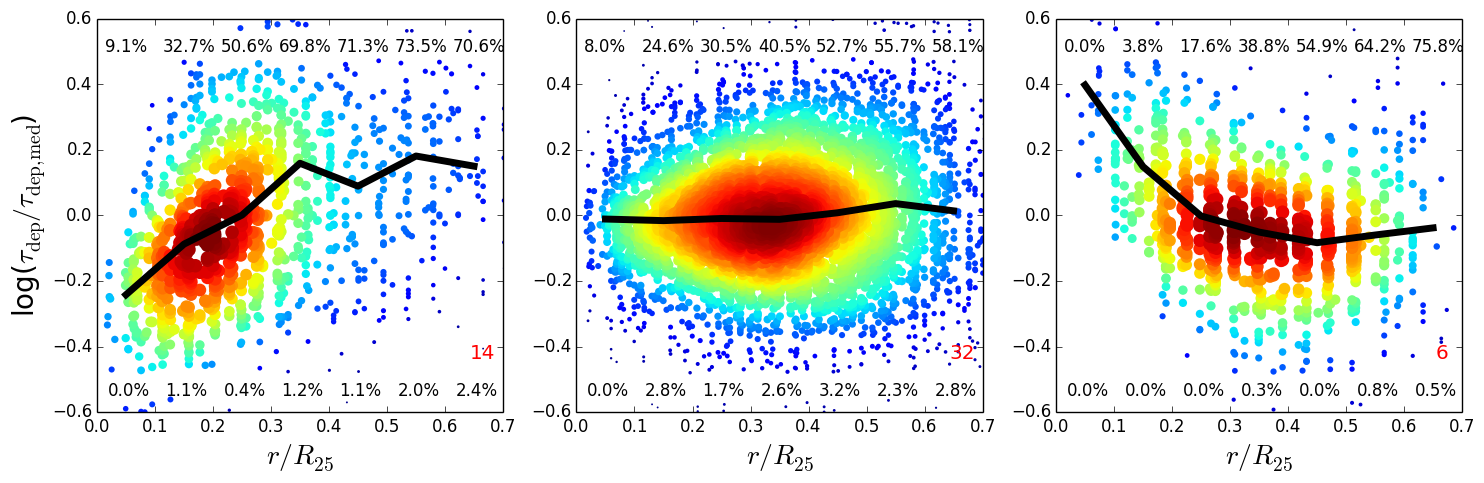}
\includegraphics[width=\textwidth]{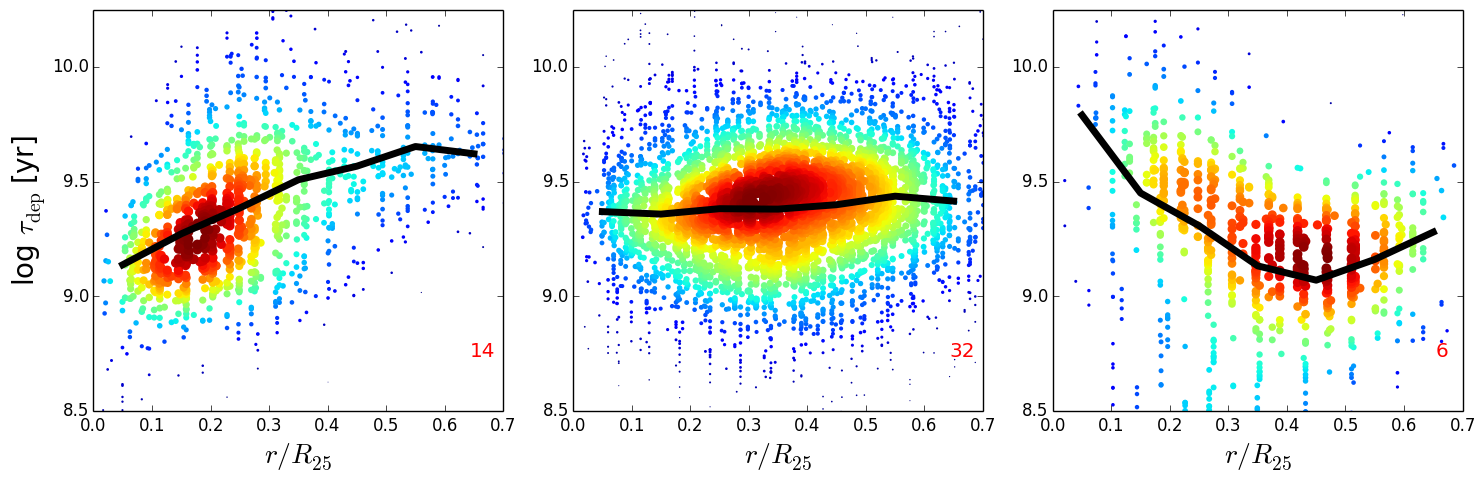}
\end{center}
\caption{Classifications of depletion time over detected regions: galaxies that show a drop of \tdep\ in the center (left), similar \tdep\ to the disk (middle), and longer \tdep\ in the center (right), in relative (top row) and absolute (bottom row) scales. The colors and sizes represent the density of data points. The median profiles for each groups are shown as black curves. The percentages on the top and bottom of top row are the fraction of non-detection, and the number of galaxies in each groups are stated in the bottom right corner of each top panels. This result extends the finding by \citet{leroy13}, where we show more complex behaviors: galactic centers can have shorter, similar, or longer \tdep\ with respect to the disk.}
\label{fig:tdep_class}
\end{figure*}

To see a clear variation of \tauctr\ with respect to \taudisk, we separate galaxies into three groups based on their log(\tauctr$/$\taudisk) values. The three groups of \tdep\ are the following. (1) Galaxies with falling \tauctr, defined as those with log$(\tau_{\rm center}/\tau_{\rm disk}) < -0.26$ dex, represent 26.9\% of the galaxy sample. (2) Galaxies with rising \tauctr, defined as those with log$(\tau_{\rm center}/\tau_{\rm disk}) > 0.26$ dex, represent 11.5\% of the galaxy sample. (3) The rest of them (61.6\% of the sample) have log($\tau_{\rm center}/\tau_{\rm disk}$) within $\pm 0.26$ dex, which we defined as flat \tdep. We list the values of \tdep\ in the centers, disks, and whole galaxy (median) in Appendix~\ref{app:gal_props}, where we use the notation "drop", "rise", and "flat" for these three groups. In this respect, we expand the previous finding that galactic centers have shorter \tdep\ than that in the disks \citep{leroy13} to include galactic centers that have similar, and even, longer \tauctr\ compared to \taudisk. The results of this segregation are shown in the top row of Figure~\ref{fig:tdep_class}.

We use 0.26 dex as a separator between 3 different groups of \tdep\ because this value is the standard deviation of resolved \tdep\ measurements within 0.7 $R_{25}$. This value also coincides with what was found in several galaxies of the HERACLES sample, which show a dip of \tauctr\ by about 0.2 dex relative to \taudisk\ \citep[for a constant CO-to-H$_2$ conversion factor;][]{leroy13}. However, keep in mind that the variation of \tauctr\ is continuous, i.e. there is no clear separation or clustering between those three groups (see Figure~\ref{fig:tdep_global}). This classification of galaxies into three groups is just an approach to see a difference between \tauctr\ and \taudisk\ in some galaxies.

We check how robust is this classification after the inclusion of upper and lower limits of \tdep\ in Appendix~\ref{sec:radial_profile}. The number of galaxies in the drop \tauctr\ group reduces from 14 to 12 after the inclusion of non-detections as $1\sigma_{\rm rms}$ and increases from 14 to 20 after the inclusion of non-detections as $2\sigma_{\rm rms}$. We refer to those numbers as the uncertainties of our classification, i.e. the number of galaxies in the drop \tauctr\ group is $14_{-2}^{+6}$. For the flat and rising \tauctr\ groups, the corresponding numbers are $32_{-4}^{+2}$ and $6_{-2}^{+0}$, respectively. About 88.5\% of the sample does not change group after the inclusion of non-detections as $1\sigma_{\rm rms}$. This means the numbers of galaxies in each group are quite robust.

In Appendix~\ref{app:res}, we check whether the drop of \tauctr\ is affected by varying physical resolutions from 1 to 3 kpc. This is equivalent to placing galaxies at farther distance. We found that the drop of \tauctr\ more prominent in a scale of 1 kpc. This means the number of galaxies in the drop \tauctr\ group is likely to be larger if we have a resolution better than 1 kpc.

In the bottom row of Figure~\ref{fig:tdep_class}, we show each three groups in the absolute scale of \tdep\ (in years). It shows that the galactic centers in the drop (rise) \tauctr\ groups form stars more (less) efficiently than those in the flat \tdep\ group, i.e. their locations in the KS diagram lie above (below) the disks. The values of \tauctr\ in the drop \tdep\ group ($\approx 1$ Gyr) are not only lower relative to \taudisk, but also in the absolute sense. Therefore, those galactic centers resemble an intermediate regime between the disks and starbursts.

\subsection{The Local Properties}

\begin{figure*}[!t]
\begin{center}
\includegraphics[width=\textwidth]{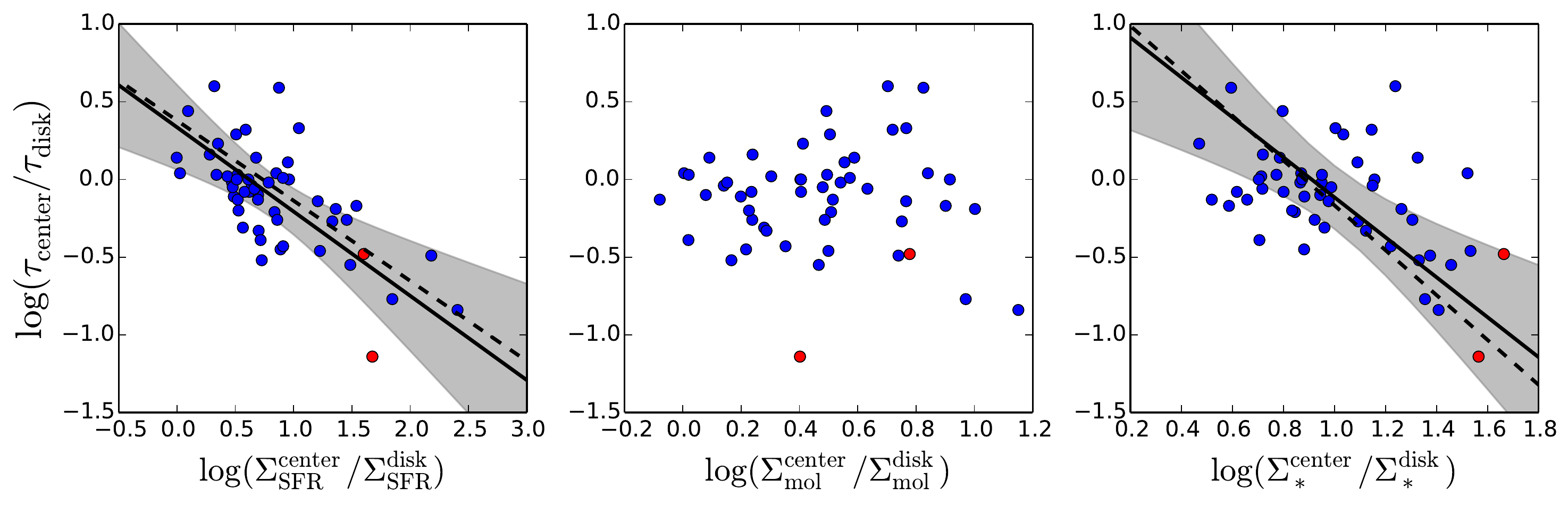}
\end{center}
\caption{The ratio between the central and disk's values for various quantities (\tdep, \sigsfr, \sigmol, and \sigstar) over detected pixels. The blue dots label late-type galaxies, while the red dots label early-type galaxies. The linear fits with equal-weight using the orthogonal distance regression \citep[ODR;][]{boggs87} in {\tt Scipy} are shown as the solid lines, while the linear fits using a likelihood-based model from \citet{kelly07} are shown as the dashed lines. The gray regions are the 95\% confidence bands from the ODR best-fit lines \citep{feigelson13}. The typical uncertainties of the data points are 0.02 dex for \sigstar, 0.1 dex for \sigmol, and 0.3 dex for \sigsfr. The slope of the correlation ($m$), the correlation coefficient ($r_c$), and the $p$-value ($p$) are tabulated in Table~\ref{tab:fig8_fit}. We do not fit the middle panels because of low $r_c$ value and high $p$ value, indicative of no correlation between log(\tauctr/\taudisk) and log$(\Sigma_{\rm mol}^{\rm center}/\Sigma_{\rm mol}^{\rm disk})$.}
\label{fig:tdep_surfden}
\end{figure*}

Is the variation of \tdep\ between the centers and the disks driven by SFR, molecular gas, or both? In Figure~\ref{fig:tdep_surfden}, we show that there is an anti-correlation between log(\tauctr/\taudisk) and log($\Sigma_{\rm SFR}^{\rm center}/\Sigma_{\rm SFR}^{\rm disk}$), but no correlation between log(\tauctr/\taudisk) and log($\Sigma_{\rm mol}^{\rm center}/\Sigma_{\rm mol}^{\rm disk}$). This means the drop of \tauctr\ is due to higher \sigsfr, not lower \sigmol\ in the center. In other words, the centers can have any values of \sigmol, but those with higher \sigsfr\ are associated with the drops of \tauctr. However, we should be cautious that the range of \sigmol\ variations ($\sim 1$ dex) is smaller than the range of \sigsfr\ variations ($\sim 2$ dex).

Then, why do some centers have higher \sigsfr, irrespective of the \sigmol\ value? In thermal and dynamical equilibrium, the weight of the ISM in the vertical gravitational field of stars and gas is balanced by the pressure created by momentum and energy from stellar feedback \citep{ostriker10,ostriker11,kim11,kim13}. Therefore, we expect a relation between \sigsfr\ (which sets the thermal, turbulent, and magnetic pressure via feedback) and \sigstar\ (which sets the ISM weight). Interestingly, in the right panels of Figure~\ref{fig:tdep_surfden} we see that log(\tauctr$/$\taudisk) correlates with the ratio of the mean values of \sigstar\ between the center and the disk. Galaxies with higher ratios of central \sigstar\ relative to those in the disks, have a drop of \tauctr. Since \sigstar\ is one of the determining factors for hydrostatic pressure \citep{blitz04,blitz06}, this means the drops of \tauctr\ are associated with high ISM pressure. Indeed, previous observations showed that the galactic center is a high pressure region \citep{spergel92,oka01,rosolowsky05}. This result suggests the star formation efficiency depends on the local environment within a galaxy.

\begin{table*}
\label{tab:fig8_fit}
\centering
\caption{The best-fit parameters from the ODR method and correlation coefficients.}
\begin{tabular}{ c c c c c c c c c }
\hline
$x-$axis & Intercept & Slope & \multicolumn{3}{c}{Correlation coefficient} & $p-$value & $p-$value & $p-$value \\
labels & $a \pm \delta a$ & $b \pm \delta b$ & Pearson & Spearman & Kendall & Pearson & Spearman & Kendall \\
\hline
SFR & $0.34 \pm 0.07$ & $-0.54 \pm 0.07$ & $-0.68$ & $-0.64$ & $-0.48$ & $3.81 \times 10^{-8}$ & $3.66 \times 10^{-7}$ & $4.29 \times 10^{-7}$ \\
Molecular & $-$ & $-$ & $-0.03$ & \phn\phn$0.04$ & \phn\phn$0.03$ & $0.84$ & $0.77$ & $0.73$ \\
Stellar & $1.17 \pm 0.22$ & $-1.29 \pm 0.22$ & $-0.49$ & $-0.39$ & $-0.27$ & $2.28 \times 10^{-4}$ & $4.02 \times 10^{-3}$ & $4.51 \times 10^{-3}$ \\
\hline
\end{tabular}
\end{table*}

\section{Discussion} \label{sec:5}

\subsection{The CO-to-H$_2$ Conversion Factor} \label{sec:alpha_co}

\begin{figure*}[!t]
\begin{center}
\includegraphics[width=\textwidth]{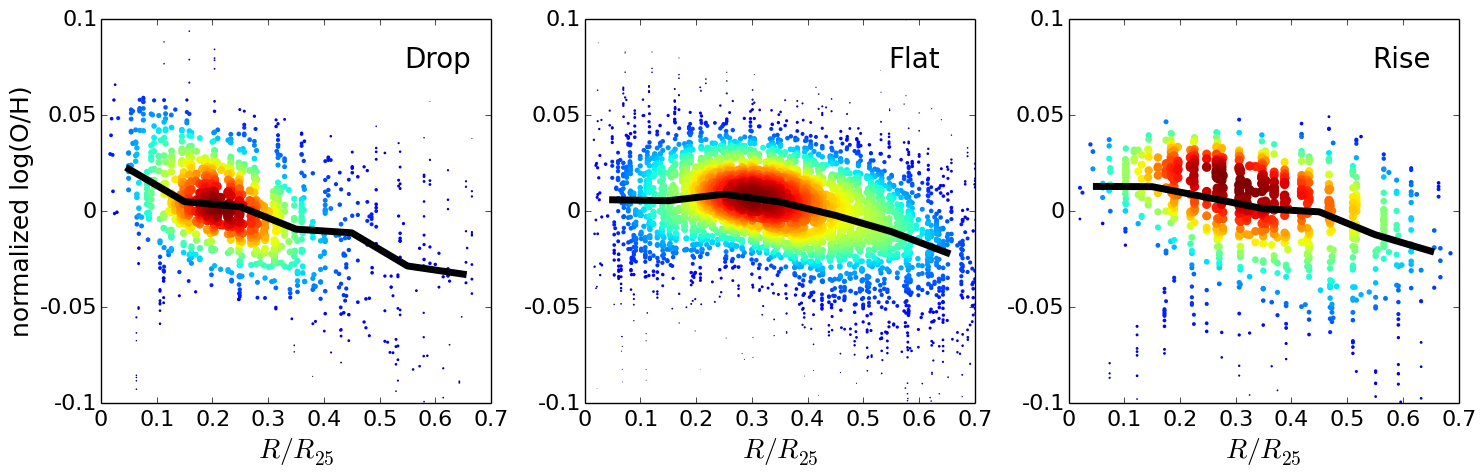}
\end{center}
\caption{Plots of the gas-phase metallicities (12+log[O/H]), relative to their median value in a galaxy, as a function of radius for the three groups: drop of \tauctr\ (left), flat \tdep\ (middle), and rise of \tauctr\ (right). The median values are shown as the solid black curves and the colors represent the density of data points. The galaxies that show drops of \tauctr\ have steeper gradient of metallicity than the other two groups.}
\label{fig:metal_class}
\end{figure*}

How is the variation of \tauctr\ affected by the change in the CO-to-H$_2$ conversion factor (\alphaco)? In general, there are two scenarios where \alphaco\ varies \citep{bolatto13}. First, the dependence of \alphaco\ with gas metallicity -- a lower gas metallicity needs a higher H$_2$ column density to shield the gas until it reaches sufficient extinction for CO to exist \citep[e.g.,][]{leroy07,leroy11}. However, the variation of metallicity from center to disk within a galaxy is very small ($\sim 0.1$ dex; Figure~\ref{fig:metal_class}), so that metallicity is unlikely to induce a significant variation on \alphaco. Furthermore, in the group that shows a drop of \tauctr, metallicities slightly rise towards the center, which means \alphaco\ is slightly lower in the center than in the disk. If we take this effect into account, it would only exaggerate the drop of \tauctr.

The second source of \alphaco\ variations is the CO emission from diffuse gas that is bound by the gravitational potential of stars and gas. Hence, the velocity dispersion of this diffuse gas ($\sigma_{\rm CO,diff}$) reflects the additional stellar gravitational potential \citep{bolatto13}. This effect increases the CO luminosity ($L_{\rm CO}$) per unit molecular gas mass because $L_{\rm CO}$ is proportional to the brightness temperature ($T_B$) and $\sigma_{\rm CO,diff}$ (assuming CO is optically thick throughout the medium). \citet{bolatto13} and \citet{sandstrom13} suggest that the variation of \alphaco\ is related to the total surface density due to stars and gas as \alphaco\ $\propto \Sigma_{\rm total}^{-\gamma}$, where $\gamma \approx 0.5$ for $\Sigma_{\rm total} > 100 \ M_{\odot}$ pc$^{-2}$. Applying this prescription for \alphaco\ would exaggerate the drop of \tauctr\ and resulting in more galaxies in the group of \tauctr\ drops.

\subsection{Metallicity Gradients}

\begin{figure*}
\begin{center}
\includegraphics[width=\textwidth]{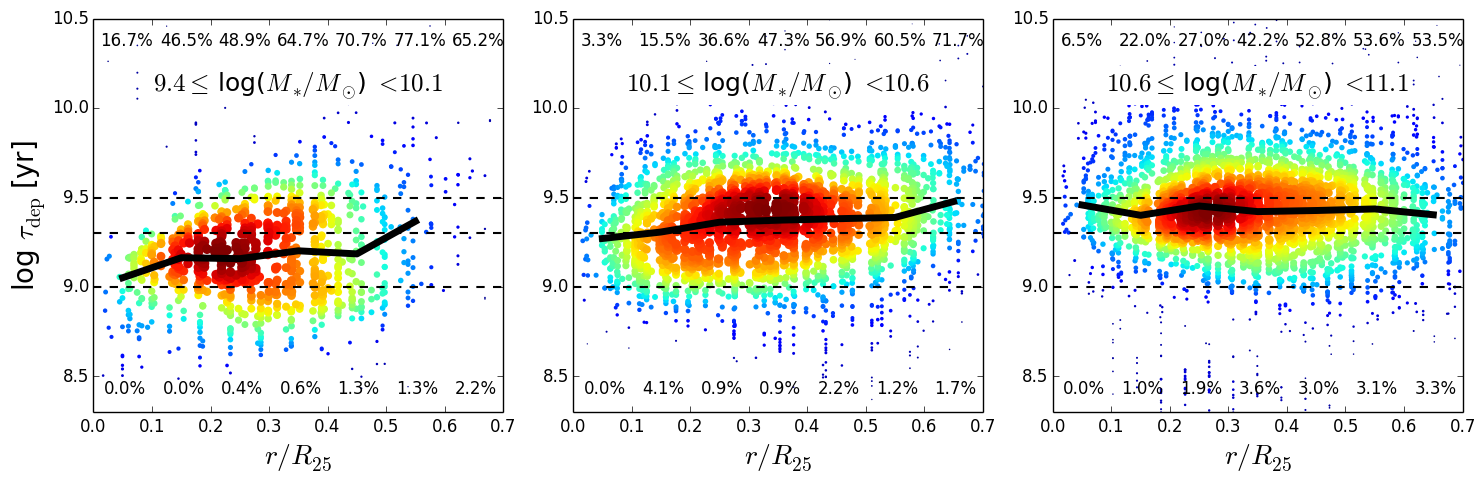}
\end{center}
\caption{The molecular gas depletion time as a function of radius, separated in three mass bins: $9.6 \leq {\rm log}(M_*/M_\odot) < 10.3$ (left panel), $10.3 \leq {\rm log}(M_*/M_\odot) < 10.8$ (middle panel), and $10.8 \leq {\rm log}(M_*/M_\odot) < 11.3$ (right panel). The colors represent the density of data points. The percentages are the fraction of upper and lower limits at a given radial bin. The solid black lines are the median value of \tdep\ at a given radial bin, while the dashed lines are the constant \tdep\ values of 1, 2, and 3 Gyrs. This figure shows that the drop of \tdep\ in the centers is more prominent in the lowest mass bin.}
\label{fig:tdep_rad_mstar}
\end{figure*}

It is interesting that the metallicity in the drop \tauctr\ group is rising toward the centers, while the metallicity profiles in the other two groups are flattening toward the centers (Figure~\ref{fig:metal_class}). In the CALIFA sample, \citet{sanchez-menguiano16} found the variation of metallicity gradients for different stellar masses: the metallicity gradient in higher mass galaxies is flattening in the center, while the metallicity gradient in lower mass galaxies is rising toward the center. Since the drop of \tdep\ is more prominent in the lowest mass bin (Figure~\ref{fig:tdep_rad_mstar}), then the variation of metallicity gradients in Figure~\ref{fig:metal_class} is possibly driven by their correlation with stellar masses. However, it remains unknown why the metallicity gradient depends on the stellar masses.

An alternative interpretation of steeper metallicity gradient is an enhancement of SFR per unit gas mass in the center (i.e. a low value of \tauctr) leads to a more metal enrichment than in the disk. Unlike stellar metallicity, gas-phase metallicity is more sensitive to the recent star formation activities, and hence, reflects the current value of \tauctr. However, the center is not a closed-box system because of inflowing gas from the disk and outflowing gas driven by the stellar feedback. Furthermore, the gas-phase metallicity is also determined by the star formation history, not only the current star formation. Therefore, the rising gradient of metallicity in the short \tauctr\ group is not clearly understood.

\subsection{The Size of the Molecular Disk}

\begin{figure*}[!ht]
\begin{center}
\includegraphics[width=\textwidth]{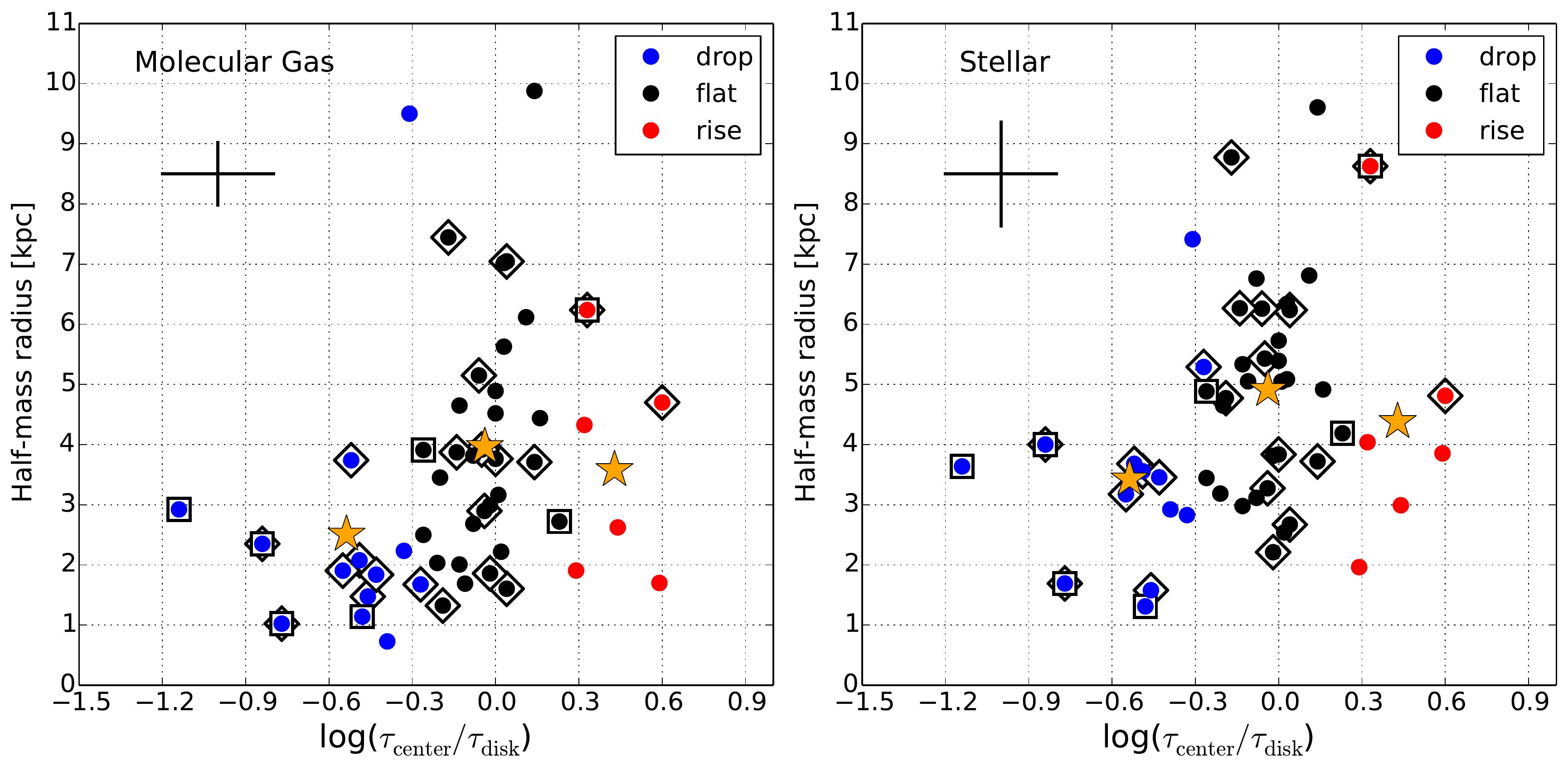}
\end{center}
\caption{The half-mass radius of molecular gas (left panel) and stars (right panel) for three groups: central drop (blue dots), flat (black dots), and central rise (red dots) of \tdep. The mean values for each three groups are marked as orange stars symbols. The typical errors are shown as crosses. The diamond symbols mark the barred galaxies, while the square symbols mark the interacting galaxies. This shows that the molecular gas distribution in the drop \tauctr\ group and in the disturbed (barred or interacting) galaxies is more compact than that in the other two groups.}
\label{fig:half_mass_rad}
\end{figure*}

In Figure \ref{fig:tdep_class}, we see that the distribution of data points in the short \tauctr\ group is more concentrated toward the center, compared to those in the flat \tdep\ group. This gives a clue that the size of the molecular disk in the short \tauctr\ group may be smaller (more compact). In order to quantify the compactness of the molecular gas and stellar distributions, we calculate the half-mass radius of molecular gas (\halfmol) and stars (\halfstar) from the cumulative distribution of \sigmol\ and \sigstar\ as a function of radius \citep{bolatto17}.

In Figure~\ref{fig:half_mass_rad}, we plot log(\tauctr$/$\taudisk) against \halfmol\ (left panel) and \halfstar\ (right panel). It turns out that galaxies in the drop \tauctr\ group have smaller \halfmol\ and \halfstar\ than those in the other two groups (quantified in Table~\ref{tab:disk_radius}). About 75\% of galaxies in the drop \tauctr\ group are disturbed systems, compared to only 44\% and 40\% for the flat and rise \tauctr\ groups, respectively. This gives a clue that the driver of physical size of the stellar and molecular gas distribution (maybe bars and interactions) is linked to the cause of \tdep\ variation in the centers. We suspect that the bar drives the gas inward toward the center (or in the case of interacting galaxies, the gas lose its angular momentum). This radial gas influx increases the pressure, resulting in higher star formation efficiency in the galactic center.

\begin{table}
\label{tab:disk_radius}
\centering
\caption{The mean molecular and stellar disk radius for each group.}
\begin{tabular}{ c c c c c }
\hline
Groups & \halfmol\ & \halfstar\ & $\bar M_*$ & $N$\tablenotemark{a} \\
{} & kpc & kpc & log$(M_\odot)$ & \\
\hline
Drop & $2.51 \pm 0.58$ & $3.43 \pm 0.41$ & $10.38 \pm 0.11$ & 14 \\
Flat & $3.96 \pm 0.34$ & $4.91 \pm 0.30$ & $10.46 \pm 0.06$ & 32 \\
Rise & $3.58 \pm 0.67$ & $4.38 \pm 0.86$ & $10.47 \pm 0.15$ & \phn 6 \\
\hline
Disturbed & $3.21 \pm 0.35$ & $4.29 \pm 0.37$ & $10.54 \pm 0.07$ & 26 \\
Undisturbed & $3.87 \pm 0.45$ & $4.63 \pm 0.35$ & $10.34 \pm 0.07$ & 26 \\
\hline
\end{tabular}
\tablenotetext{a}{The number of galaxies in each group.}
\tablecomments{The uncertainty is calculated from the standard deviation divided by the square-root of the number of galaxies.}
\end{table}

\subsection{A Burst of Star Formation} \label{sec:burst}

\begin{figure*}
\begin{center}
\includegraphics[width=\textwidth]{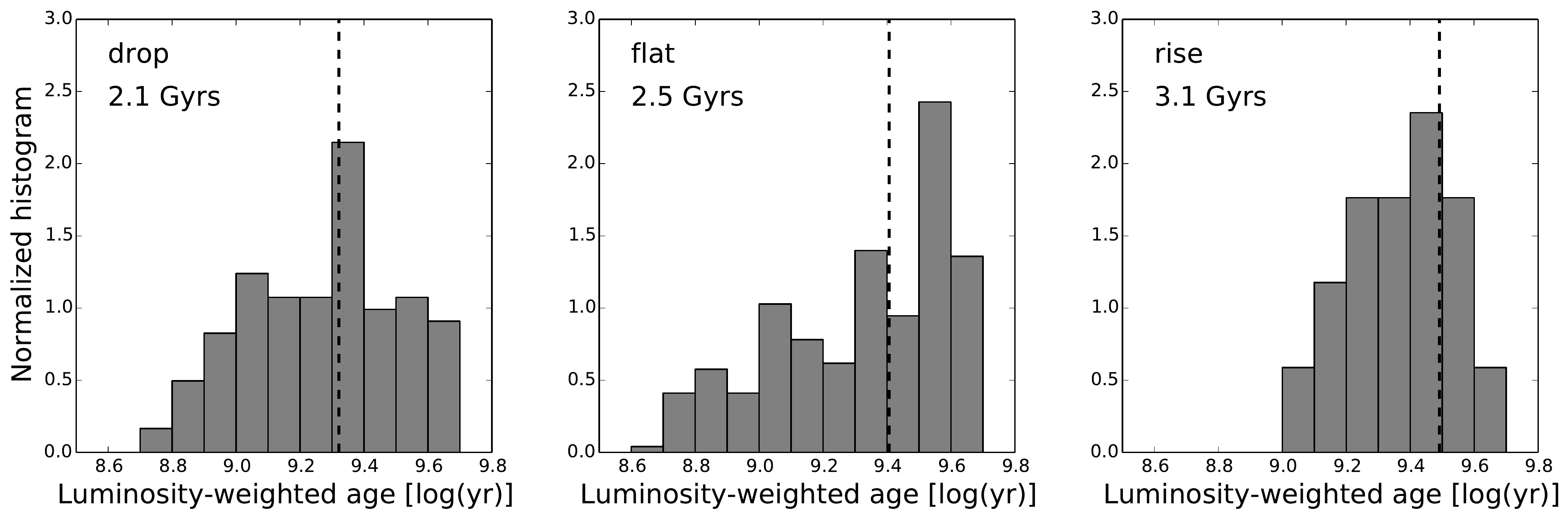}
\end{center}
\caption{The histogram of luminosity-weighted stellar population ages \citep[from CALIFA;][]{sanchez16} in the center of galaxies ($r \leq 0.1 \ R_{25}$) that show a drop of \tauctr\ (left panel), flat \tdep\ (middle panel), and a rise of \tauctr\ (right panel). The dashed lines mark the median ages, with their values are noted in the top left corner of each panel. The stellar populations in the centers of drop \tauctr\ group tend to be younger than the other two groups, consistent with the idea of a bursting period of star formation.}
\label{fig:tdep_age}
\end{figure*}

For galaxies in the drop \tauctr\ group, there may be a central starburst activity on scales below our resolution as indicated by the stellar population ages. There are at least two tracers of the stellar population ages: the UV-to-H$\alpha$ ratio \citep[e.g.,][]{leroy12,weisz12} and the age derived from the stellar population synthesis \citep[which is available in the IFU data products of][]{sanchez16}. Since we do not have the resolved UV maps in hand, we rely on the second tracer. In Figure~\ref{fig:tdep_age}, we show the histogram of the luminosity-weighted ages of stellar populations in the centers ($r < 0.1 \ R_{25}$) for each \tdep\ group. It turns out that the centers in the drop \tauctr\ group (left panel) tend to have younger ages of stellar populations ($\approx 2.1 \pm 1.1$ Gyrs) than the other two groups ($\approx 2.5 \pm 1.6$ Gyrs and $\approx 3.1 \pm 1.6$ Gyrs; middle and right panels).

We do a Kolmogorov-Smirnov  test to check whether the age distributions in each group can be drawn from the same underlying distribution. The $p-$values between the age distributions in the central drop of \tdep\ and the other two groups are $2 \times 10^{-5}$ and $0.07$, while the $p-$value between the flat and rise \tdep\ group is $0.49$. A small $p-$value means the distributions of the two samples are distinct. An Anderson-Darling test to those distributions also yields similar results: the $p-$values between the drop \tauctr\ group and the other two groups are $9 \times 10^{-5}$ and 0.02, while the $p-$value between the flat and rise groups is 0.61. This evidence strengthens our suspicion that the centers of the short \tauctr\ group are currently undergoing a burst of star formation. However, further high resolution data are needed to confirm this hypothesis.

\section{Summary} \label{sec:6}

We present results from the EDGE survey, a first major, resolved CO follow-up to an IFU survey of local galaxies (CALIFA). We combine the CO and optical IFU data to study the variation of \tdep\ between the centers and the disks in 52 local galaxies. Our findings are the following.

\begin{enumerate}

\item Contrary to the well-defined value of \tdep\ in galactic disks, galactic centers can have shorter, longer, or similar \tdep\ compared to their disks (Figure~\ref{fig:tdep_class}). The short \tauctr\ group (representing 26.9\% of the samples with \tauctr\ $\sim 1$ Gyr) resembles the intermediate regime between the disks (\taudisk\ $\sim 2.4$ Gyrs) and starbursts (\tdep\ $\sim 0.2$ Gyrs). Applying the variations of CO-to-H$_2$ conversion factor (that depends on the total surface density and metallicities) only exaggerates the drop of \tauctr.

\item The drop of \tauctr\ is caused by higher central \sigsfr\ than those in the disk, not lower \sigmol\ (Figure~\ref{fig:tdep_surfden}). Furthermore, galaxies with the higher contrast of stellar surface density in the center (i.e. higher $\Sigma_*^{\rm center}/\Sigma_*^{\rm disk}$) tend to have shorter \tauctr/\taudisk. Since the dynamical equilibrium pressure depends on \sigstar\ \citep{blitz04,blitz06,ostriker10}, this suggests that the central drop in \tdep\ is driven by high gas pressure.  This is expected for the star formation self-regulated model, in which the star formation rate locally adjusts so that feedback from massive stars offsets turbulent energy dissipation and cooling. A high feedback rate (short \tdep) is required to maintain the high pressure in regions where the vertical gravity from stars and gas is very strong \citep{ostriker10,ostriker11,kim11,kim13}.

\item The gradient of oxygen abundance rises toward the center for galaxies in the short \tauctr\ group, while the gradient is flat in the center of other groups (Figure~\ref{fig:metal_class}). This could be the stellar mass effect, where the gradient of oxygen abundance is flattening in massive galaxies \citep[as found by][]{sanchez-menguiano16}, or the oxygen abundance is sensitive to the current star formation efficiency. However, the narrow range of the oxygen abundance variation in our sample ($\sim 0.2$ dex) becomes the limitation of our analysis.

\item There are two signatures for dynamical effect that drives the variation of \tauctr\ versus \taudisk. First, the barred and interacting galaxies tend to have lower values of log(\tauctr$/$\taudisk) than the unbarred, isolated galaxies (Figure~\ref{fig:tdep_global}). Second, the size of molecular gas disk is smaller in the drop \tauctr\ group than in the other groups (Figure~\ref{fig:half_mass_rad}). We suspect that the bar drives the gas inward toward the center (or in the case of interacting galaxies, the gas lose its angular momentum). This radial gas compression increase the pressure, and resulting in higher star formation efficiency in the galactic center \citep{krumholz15}.

\end{enumerate}

In conclusion, these findings imply that the formation of stars from the molecular gas depends on the local environment within a galaxy (such as \sigstar) and the galaxy dynamics induced by bar or interactions. In the future, we are interested to measure the dense gas (as traced by HCN lines) to investigate whether the short \tauctr\ is also due to a higher fraction of the dense gas in the center. In addition, measuring the shear rate and the inflow speed in barred galaxies will give a better evidence of the importance of galactic dynamics in driving \tdep. Finally, expanding our sample towards early-type and low mass galaxies using ALMA is a natural approach to expand our statistical sample in the three groups of \tdep.

\acknowledgments
We thank the referee, Christine Wilson, for her valuable inputs that greatly improved the manuscript. We also thank John Carpenter for his help in managing the schedule of CARMA observations, and Chris McKee for insightful discussion.

The works of DU and LB are supported by the National Science Foundation (NSF) under grants AST-1140063 and AST-1616924. ADB and RCL acknowledge support from NSF through grants AST-1412419 and AST-1615960. ADB also acknowledges visiting support by the Alexander von Humboldt Foundation. TW and YC acknowledge support from NSF through grants AST-1139950 and AST-1616199. The work of ECO is supported by the NSF under grant AST-1312006. SFS acknowledges the PAPIIT-DGAPA-IA101217 project and CONACYT-IA-180125. RGB acknowledges support through grant AYA2016-77846-P. ER is supported by a Discovery Grant from NSERC of Canada. SV acknowledges support from NSF AST-1615960.

We acknowledge the usage of the HyperLeda database (\href{<url>}{http://leda.univ-lyon1.fr}). Support for the CARMA construction was derived from the states of California, Illinois, and Maryland, the James S. McDonnell Foundation, the Gordon and Betty Moore Foundation, the Kenneth T. and Eileen L. Norris Foundation, the University of Chicago, the Associates of the California Institute of Technology, and NSF. This research is based on observations collected at the Centro Astronomico Hispano Aleman (CAHA) at Calar Alto, operated jointly by the Max-Planck Institute for Astronomy (MPIA) and the Instituto de Astrofisica de Andalucia (CSIC).

\software{Pipe3D \citep[version 2.2;][]{sanchez16}, MIRIAD \citep{sault95}}, idl\textunderscore mommaps.pro \citep{wong13}, linmix\textunderscore err.pro \citep{kelly07}, matplotlib \citep{hunter07}, and SciPy \citep{jones01}.

\appendix

\section{List of galaxy properties in the sample} \label{app:gal_props}

\begin{deluxetable*}{ l l c c r r r c r c r c c c c c c }
\label{tab:samples}
\tablewidth{0pt}
\tabletypesize{\scriptsize}
\tablecaption{The list of galaxy properties in the sample.}
\tablehead{
\colhead{No.} & Galaxies & RA & Dec & $M_*$\tablenotemark{a} & $M_{\rm mol}$\tablenotemark{b} & $R_{25}$\tablenotemark{c} & Beam\tablenotemark{d} & Dist.\tablenotemark{e} & Inc.\tablenotemark{f} & P.A.\tablenotemark{f} & $\tau_{\rm center}$ & \taudisk
 & $\tau_{\rm med}$ & Group & Bar\tablenotemark{g} & Inter.\tablenotemark{h}\\
{} & {} & h:m:s & d:m:s & \multicolumn{2}{c}{log($M_\odot$)} & kpc & kpc & Mpc & deg. & deg. & log(yr) & log(yr) & log(yr)
}
\startdata
 1 & IC1151 & $15^{\rm h} 58^{\rm m} 58\fs5$ & $17^\circ 26\arcmin 26\farcs 5$ & 9.82 & 7.93 & 10.01 & 0.67 & 30.80 & 68.0 & 208.9 & 8.94 & 9.04 & 8.99 & flat & N & N \\
 2 & IC1199 & $16^{\rm h} 10^{\rm m} 10\fs6$ & $10^\circ 02\arcmin 02\farcs 4$ & 10.58 & 9.35 & 11.83 & 1.52 & 68.25 & 64.5 & 337.3 & 9.45 & 9.58 & 9.56 & flat & N & N \\
 3 & IC1683 & $01^{\rm h} 22^{\rm m} 22\fs6$ & $34^\circ 26\arcmin 26\farcs 2$ & 10.56 & 9.68 & 13.34 & 1.47 & 69.73 & 44.8 & 20.6 & 9.15 & 9.64 & 9.64 & drop & Y & N \\
 4 & NGC0477 & $01^{\rm h} 21^{\rm m} 21\fs3$ & $40^\circ 29\arcmin 29\farcs 3$ & 10.70 & 9.54 & 19.29 & 1.86 & 85.42 & 60.0 & 150.0 & 9.37 & 9.68 & 9.66 & drop & N & N \\
 5 & NGC0496 & $01^{\rm h} 23^{\rm m} 23\fs2$ & $33^\circ 31\arcmin 31\farcs 7$ & 10.64 & 9.48 & 11.34 & 1.82 & 87.47 & 57.0 & 38.5 & 9.15 & 9.23 & 9.22 & flat & N & N \\
 6 & NGC0551 & $01^{\rm h} 27^{\rm m} 27\fs7$ & $37^\circ 10\arcmin 11\farcs 0$ & 10.75 & 9.39 & 16.10 & 1.54 & 74.50 & 64.2 & 320.0 & 9.62 & 9.58 & 9.61 & flat & Y & N \\
 7 & NGC2253 & $06^{\rm h} 43^{\rm m} 43\fs7$ & $65^\circ 12\arcmin 12\farcs 4$ & 10.60 & 9.62 & 10.61 & 1.20 & 51.16 & 47.4 & 300.0 & 9.37 & 9.37 & 9.37 & flat & Y & N \\
 8 & NGC2347 & $07^{\rm h} 16^{\rm m} 16\fs1$ & $64^\circ 42\arcmin 42\farcs 6$ & 10.84 & 9.56 & 15.25 & 1.49 & 63.75 & 50.2 & 189.1 & 9.48 & 9.34 & 9.38 & flat & Y & N \\
 9 & NGC2730 & $09^{\rm h} 02^{\rm m} 02\fs3$ & $16^\circ 50\arcmin 50\farcs 3$ & 9.93 & 9.00 & 11.52 & 1.26 & 54.78 & 27.7 & 260.8 & 9.13 & 9.24 & 9.23 & flat & N & N \\
10 & NGC2906 & $09^{\rm h} 32^{\rm m} 32\fs1$ & $08^\circ 26\arcmin 26\farcs 5$ & 10.38 & 9.11 & 7.44 & 0.94 & 37.73 & 55.7 & 265.0 & 9.78 & 9.34 & 9.40 & rise & N & N \\
11 & NGC3381 & $10^{\rm h} 48^{\rm m} 48\fs4$ & $34^\circ 42\arcmin 42\farcs 7$ & 9.68 & 8.11 & 6.87 & 0.50 & 23.40 & 30.8 & 43.1 & 8.86 & 9.31 & 9.30 & drop & Y & N \\
12 & NGC3811 & $11^{\rm h} 41^{\rm m} 41\fs3$ & $47^\circ 41\arcmin 41\farcs 4$ & 10.44 & 9.28 & 13.05 & 0.96 & 44.25 & 42.5 & 359.0 & 9.32 & 9.28 & 9.31 & flat & Y & N \\
13 & NGC3815 & $11^{\rm h} 41^{\rm m} 41\fs7$ & $24^\circ 48\arcmin 48\farcs 0$ & 10.32 & 9.16 & 11.22 & 1.14 & 53.59 & 59.9 & 67.8 & 9.43 & 9.47 & 9.45 & flat & Y & N \\
14 & NGC3994 & $11^{\rm h} 57^{\rm m} 57\fs6$ & $32^\circ 16\arcmin 16\farcs 6$ & 10.39 & 9.26 & 5.53 & 1.02 & 44.75 & 59.5 & 188.1 & 9.07 & 8.78 & 8.81 & rise & N & N \\
15 & NGC4047 & $12^{\rm h} 02^{\rm m} 02\fs8$ & $48^\circ 38\arcmin 38\farcs 2$ & 10.67 & 9.66 & 10.95 & 1.06 & 49.06 & 42.1 & 105.0 & 9.41 & 9.43 & 9.41 & flat & N & N \\
16 & NGC4470 & $12^{\rm h} 29^{\rm m} 29\fs6$ & $07^\circ 49\arcmin 49\farcs 4$ & 10.03 & 8.59 & 6.23 & 0.78 & 33.43 & 47.5 & 359.5 & 8.74 & 8.87 & 8.85 & flat & N & N \\
17 & NGC4644 & $12^{\rm h} 42^{\rm m} 42\fs7$ & $55^\circ 08\arcmin 08\farcs 7$ & 10.48 & 9.20 & 15.77 & 1.60 & 71.65 & 72.9 & 57.0 & 9.59 & 9.56 & 9.57 & flat & N & N \\
18 & NGC4711 & $12^{\rm h} 48^{\rm m} 48\fs8$ & $35^\circ 19\arcmin 20\farcs 0$ & 10.38 & 9.18 & 10.31 & 1.32 & 58.83 & 58.3 & 215.0 & 9.60 & 9.44 & 9.45 & flat & N & N \\
19 & NGC4961 & $13^{\rm h} 05^{\rm m} 05\fs8$ & $27^\circ 44\arcmin 44\farcs 0$ & 9.77 & 8.41 & 5.93 & 0.78 & 36.58 & 46.6 & 90.0 & 9.21 & 9.23 & 9.22 & flat & Y & N \\
20 & NGC5000 & $13^{\rm h} 09^{\rm m} 09\fs8$ & $28^\circ 54\arcmin 54\farcs 4$ & 10.74 & 9.45 & 15.04 & 1.62 & 80.80 & 20.0 & 1.3 & 9.40 & 9.59 & 9.53 & flat & Y & N \\
21 & NGC5016 & $13^{\rm h} 12^{\rm m} 12\fs1$ & $24^\circ 05\arcmin 05\farcs 7$ & 10.27 & 8.90 & 8.45 & 0.83 & 36.90 & 39.9 & 57.4 & 9.10 & 9.43 & 9.40 & drop & N & N \\
22 & NGC5056 & $13^{\rm h} 16^{\rm m} 16\fs2$ & $30^\circ 57\arcmin 57\farcs 0$ & 10.64 & 9.45 & 19.14 & 1.96 & 81.14 & 61.4 & 178.0 & 9.03 & 8.43 & 8.51 & rise & Y & N \\
23 & NGC5480 & $14^{\rm h} 06^{\rm m} 06\fs4$ & $50^\circ 43\arcmin 43\farcs 5$ & 9.97 & 8.92 & 6.57 & 0.52 & 26.96 & 41.5 & 178.0 & 8.99 & 9.20 & 9.20 & flat & N & N \\
24 & NGC5520 & $14^{\rm h} 12^{\rm m} 12\fs4$ & $50^\circ 20\arcmin 20\farcs 9$ & 9.87 & 8.67 & 6.25 & 0.55 & 26.73 & 59.1 & 245.1 & 8.99 & 9.45 & 9.30 & drop & Y & N \\
25 & NGC5633 & $14^{\rm h} 27^{\rm m} 27\fs5$ & $46^\circ 08\arcmin 08\farcs 8$ & 10.20 & 9.14 & 5.29 & 0.71 & 33.38 & 41.9 & 16.9 & 9.25 & 9.23 & 9.24 & flat & N & N \\
26 & NGC5657 & $14^{\rm h} 30^{\rm m} 30\fs7$ & $29^\circ 10\arcmin 10\farcs 8$ & 10.29 & 9.11 & 14.34 & 1.20 & 56.33 & 68.3 & 344.0 & 9.00 & 9.55 & 9.52 & drop & Y & N \\
27 & NGC5732 & $14^{\rm h} 40^{\rm m} 40\fs7$ & $38^\circ 38\arcmin 38\farcs 3$ & 10.03 & 8.82 & 9.66 & 1.25 & 54.00 & 58.4 & 43.2 & 9.16 & 9.42 & 9.41 & flat & N & N \\
28 & NGC5784 & $14^{\rm h} 54^{\rm m} 54\fs3$ & $42^\circ 33\arcmin 33\farcs 5$ & 11.09 & 9.40 & 17.12 & 1.67 & 79.42 & 45.0 & 252.0 & 9.26 & 10.40 & 9.95 & drop & N & Y \\
29 & NGC5930 & $15^{\rm h} 26^{\rm m} 26\fs1$ & $41^\circ 40\arcmin 40\farcs 6$ & 10.40 & 9.33 & 10.01 & 0.83 & 37.23 & 45.0 & 155.0 & 9.27 & 10.04 & 9.71 & drop & Y & Y \\
30 & NGC5934 & $15^{\rm h} 28^{\rm m} 28\fs2$ & $42^\circ 55\arcmin 55\farcs 8$ & 10.66 & 9.81 & 7.35 & 1.76 & 82.71 & 55.0 & 5.0 & 10.00 & 9.77 & 9.79 & flat & N & Y \\
31 & NGC5947 & $15^{\rm h} 30^{\rm m} 30\fs6$ & $42^\circ 43\arcmin 43\farcs 0$ & 10.67 & 9.26 & 14.61 & 1.92 & 86.07 & 32.2 & 206.6 & 9.09 & 9.61 & 9.59 & drop & Y & N \\
32 & NGC5953 & $15^{\rm h} 34^{\rm m} 34\fs5$ & $15^\circ 11\arcmin 11\farcs 6$ & 10.18 & 9.49 & 6.09 & 0.61 & 28.43 & 26.1 & 43.3 & 9.12 & 9.60 & 9.47 & drop & N & Y \\
33 & NGC5980 & $15^{\rm h} 41^{\rm m} 41\fs5$ & $15^\circ 47\arcmin 47\farcs 3$ & 10.61 & 9.70 & 14.10 & 1.27 & 59.36 & 66.2 & 15.0 & 9.47 & 9.15 & 9.19 & rise & N & N \\
34 & NGC6004 & $15^{\rm h} 50^{\rm m} 50\fs4$ & $18^\circ 56\arcmin 56\farcs 4$ & 10.66 & 9.33 & 15.19 & 1.22 & 55.21 & 37.3 & 277.3 & 9.61 & 9.66 & 9.63 & flat & Y & N \\
35 & NGC6060 & $16^{\rm h} 05^{\rm m} 05\fs9$ & $21^\circ 29\arcmin 29\farcs 1$ & 10.78 & 9.68 & 17.41 & 1.28 & 63.24 & 64.3 & 102.0 & 9.39 & 9.36 & 9.38 & flat & N & N \\
36 & NGC6155 & $16^{\rm h} 26^{\rm m} 26\fs1$ & $48^\circ 22\arcmin 22\farcs 0$ & 10.18 & 8.94 & 6.68 & 0.77 & 34.60 & 44.7 & 130.0 & 9.02 & 9.10 & 9.08 & flat & N & N \\
37 & NGC6186 & $16^{\rm h} 34^{\rm m} 34\fs4$ & $21^\circ 32\arcmin 32\farcs 5$ & 10.41 & 9.46 & 9.68 & 0.92 & 42.38 & 71.2 & 69.8 & 9.32 & 9.49 & 9.46 & flat & Y & N \\
38 & NGC6301 & $17^{\rm h} 08^{\rm m} 08\fs5$ & $42^\circ 20\arcmin 20\farcs 3$ & 10.98 & 9.96 & 31.24 & 2.63 & 121.36 & 52.8 & 288.5 & 9.75 & 9.61 & 9.65 & flat & N & N \\
39 & NGC7738 & $23^{\rm h} 44^{\rm m} 44\fs0$ & $00^\circ 31\arcmin 31\farcs 0$ & 11.01 & 9.99 & 16.87 & 1.90 & 97.82 & 65.6 & 244.7 & 9.17 & 10.01 & 9.74 & drop & Y & Y \\
40 & NGC7819 & $00^{\rm h} 04^{\rm m} 04\fs4$ & $31^\circ 28\arcmin 28\farcs 3$ & 10.41 & 9.27 & 14.99 & 1.43 & 71.62 & 54.0 & 280.3 & 9.28 & 9.55 & 9.54 & drop & Y & N \\
41 & UGC03253 & $05^{\rm h} 19^{\rm m} 19\fs7$ & $84^\circ 03\arcmin 03\farcs 1$ & 10.43 & 8.88 & 11.88 & 1.57 & 59.46 & 58.3 & 267.7 & 8.88 & 9.31 & 9.29 & drop & Y & N \\
42 & UGC04132 & $07^{\rm h} 59^{\rm m} 59\fs2$ & $32^\circ 54\arcmin 54\farcs 9$ & 10.74 & 10.02 & 13.51 & 1.70 & 75.35 & 72.0 & 212.6 & 9.35 & 9.41 & 9.41 & flat & Y & N \\
43 & UGC04461 & $08^{\rm h} 33^{\rm m} 33\fs4$ & $52^\circ 31\arcmin 31\farcs 9$ & 10.17 & 9.24 & 14.51 & 1.59 & 72.27 & 70.1 & 215.8 & 9.36 & 9.35 & 9.36 & flat & N & N \\
44 & UGC05108 & $09^{\rm h} 35^{\rm m} 35\fs4$ & $29^\circ 48\arcmin 48\farcs 8$ & 10.90 & 9.75 & 18.84 & 2.81 & 118.41 & 66.1 & 133.1 & 9.47 & 9.61 & 9.53 & flat & Y & N \\
45 & UGC07012 & $12^{\rm h} 02^{\rm m} 02\fs1$ & $29^\circ 50\arcmin 50\farcs 9$ & 9.70 & 8.35 & 6.96 & 0.92 & 44.28 & 60.5 & 182.1 & 8.75 & 9.14 & 9.13 & drop & N & N \\
46 & UGC08107 & $12^{\rm h} 59^{\rm m} 59\fs7$ & $53^\circ 20\arcmin 20\farcs 5$ & 11.00 & 10.11 & 40.43 & 2.75 & 121.62 & 71.4 & 233.2 & 9.92 & 9.59 & 9.60 & rise & Y & Y \\
47 & UGC09067 & $14^{\rm h} 10^{\rm m} 10\fs8$ & $15^\circ 12\arcmin 12\farcs 6$ & 10.76 & 9.83 & 13.54 & 2.75 & 114.50 & 62.4 & 14.6 & 9.46 & 9.46 & 9.46 & flat & N & N \\
48 & UGC09476 & $14^{\rm h} 41^{\rm m} 41\fs5$ & $44^\circ 30\arcmin 30\farcs 8$ & 10.23 & 9.15 & 10.19 & 1.01 & 46.63 & 48.5 & 312.0 & 9.32 & 9.52 & 9.50 & flat & N & N \\
49 & UGC09542 & $14^{\rm h} 49^{\rm m} 49\fs0$ & $42^\circ 27\arcmin 27\farcs 8$ & 10.32 & 9.31 & 16.64 & 1.65 & 79.70 & 72.7 & 214.3 & 9.56 & 9.56 & 9.56 & flat & N & N \\
50 & UGC09759 & $15^{\rm h} 10^{\rm m} 10\fs7$ & $55^\circ 21\arcmin 21\farcs 0$ & 9.81 & 9.07 & 9.55 & 1.03 & 49.25 & 66.8 & 54.7 & 10.20 & 9.61 & 9.69 & rise & N & N \\
51 & UGC10205 & $16^{\rm h} 06^{\rm m} 06\fs7$ & $30^\circ 05\arcmin 05\farcs 9$ & 10.88 & 9.60 & 19.95 & 2.21 & 94.92 & 51.7 & 118.6 & 9.56 & 9.82 & 9.81 & flat & N & Y \\
52 & UGC10710 & $17^{\rm h} 06^{\rm m} 06\fs9$ & $43^\circ 07\arcmin 07\farcs 3$ & 10.72 & 9.88 & 29.51 & 2.63 & 121.69 & 69.6 & 329.5 & 9.61 & 9.50 & 9.50 & flat & N & N \\
\enddata
\tablenotetext{a}{The stellar mass assuming Kroupa IMF from the CALIFA survey \citep{sanchez16}.}
\tablenotetext{b}{The molecular gas mass assuming CO-to-H$_2$ conversion factor of 4.4 $M_\odot$ pc$^{-2}$ (K km s$^{-1}$ pc$^2$)$^{-1}$ from the EDGE survey \citep{bolatto17}, including mass contribution from Helium.}
\tablenotetext{c}{The radius where the surface brightness is 25 mag arcsec$^{-2}$ in the $B-$band, from the HyperLEDA catalog \citep{makarov14}.}
\tablenotetext{d}{The physical beam size, calculated from the geometric mean of the major and minor axes of the EDGE beam.}
\tablenotetext{e}{The luminosity distance computed from the CALIFA redshift for ionized gas lines assuming $H_0 = 70$ km s$^{-1}$, $\Omega_{\rm m} = 0.27$, and $\Omega_{\Lambda}=0.73$.}
\tablenotetext{f}{The inclination and position angle are taken from the following, ordered by priority: (1) the best fit of CO rotation curve (Levy et al. in preparation), whenever it is possible, (2) from the shape of the outer isophote, or (3) from the HyperLEDA catalog \citep{makarov14}.}
\tablenotetext{g}{The bar assignments ({\tt Yes} or {\tt No}) are taken from the following, ordered by priority: (1) the photometric fit from \citet{mendez-abreu17}, or (2) the HyperLEDA catalog \citep{makarov14}.}
\tablenotetext{h}{The assignment for interacting galaxies ({\tt Yes} or {\tt No}), taken from \citet{barrera-ballesteros15}.}
\end{deluxetable*}

\section{The Effect of Non-detections} \label{sec:radial_profile}

\begin{figure*}
\begin{center}
\includegraphics[width=\textwidth]{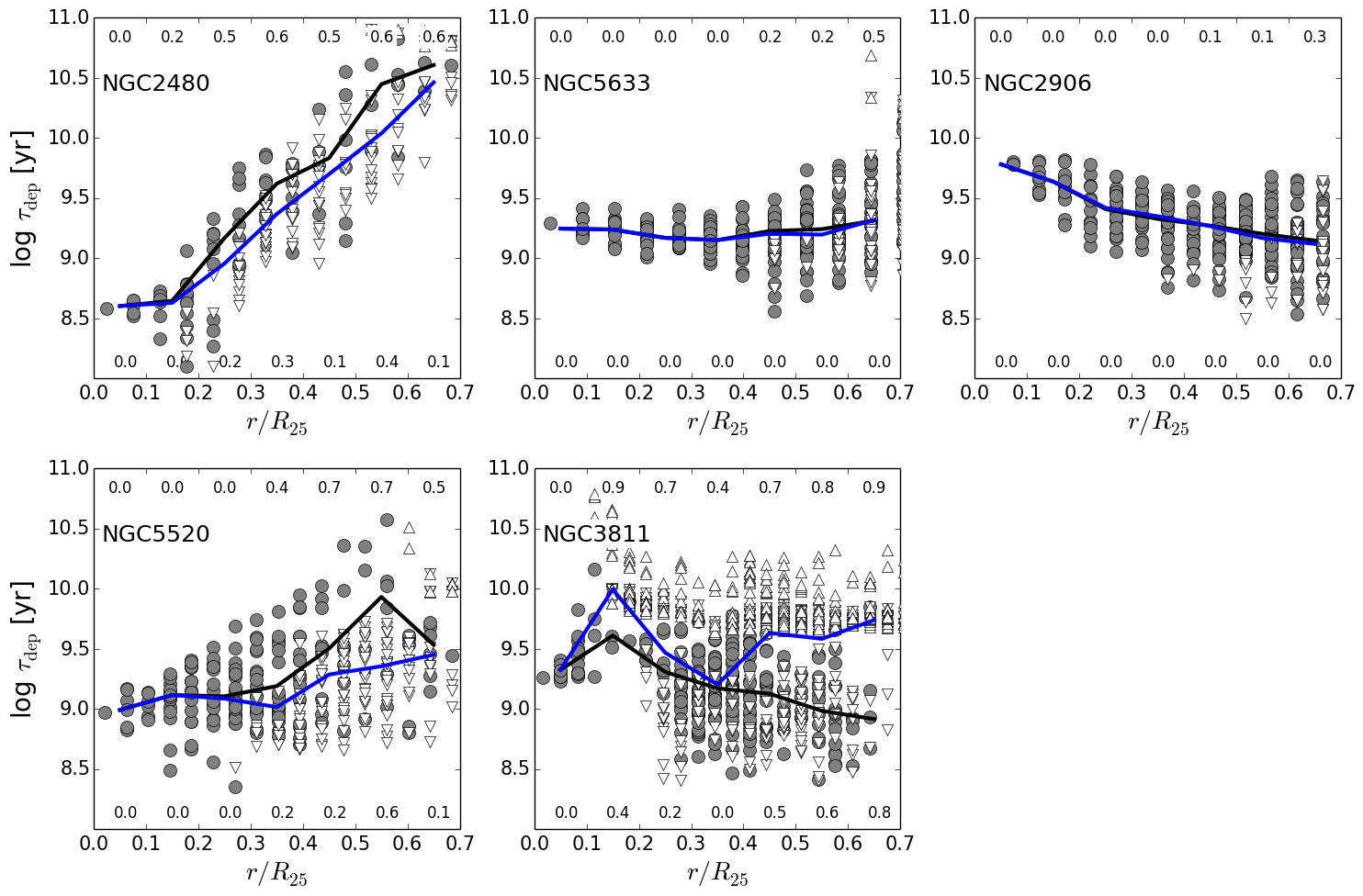}
\end{center}
\caption{Depletion time as a function of radius for three groups: drop (left panel), flat (middle panel), and rising (right panel) \tdep\ in the center. The gray circles are the detection points. The upper and lower limits are marked as the triangles pointing downward and upward, respectively. The black lines are the median of the local \tdep\ in radial bins based on detections only, while the blue lines are the median of the local \tdep\ in radial bins after the inclusion of non-detections as $1 \sigma_{\rm rms}$. Each panel in the {\it top row} is a galaxy that has the {\it same classification} in both the detection only and after the inclusion of non-detections as $1 \sigma_{\rm rms}$. Each panel in the {\it bottom row} shows a galaxy that is classified as drop (left panel) and flat (middle panel) categories based on the detection only, but is classified as flat and rise, respectively, after the inclusion of non-detections as $1 \sigma_{\rm rms}$ (see Table~\ref{tab:compare_class2}). The decimal numbers on the top and bottom of each panel are the fractions of non-detections and the absolute difference (in dex) between the black and blue lines, respectively.}
\label{fig:tdep_prof_class}
\end{figure*}

The classification of \tauctr\ in $\S$\ref{sec:classify} only takes into account the detected regions in both \sigmol\ and \sigsfr\ (shown as gray circles in Figure~\ref{fig:tdep_prof_class}). We now check the robustness of our results by including the upper and lower limits of \tdep. For the upper limit of \tdep, \sigmol\ is non-detected and is replaced by $1\sigma_{\rm rms}$, while \sigsfr\ is detected. Conversely, for the lower limit of \tdep, \sigmol\ is detected, while \sigsfr\ is not-detected and is replaced by $1\sigma_{\rm rms}$. The upper and lower limits of \tdep\ are shown as triangles pointing down and up, respectively, in Figure~\ref{fig:tdep_prof_class}. Then, we calculate the median value of \tdep\ (after the inclusion of upper and lower limits) in each radial bin (shown as the blue lines in Figure~\ref{fig:tdep_prof_class}). As a comparison, the median values of \tdep\ using only the detected regions in radial bins are shown as the black lines. The upper limits tend to have lower \tdep\ than that in detected regions. Therefore, the blue line can be lower than the black line where upper limits are dominant (as in NGC2480 and NGC5520). Inverse situation happens where lower limits are dominant (as in NGC3811). If detected regions are dominant then the blue and black lines are coincidence with each other (as in NGC5633 and NGC2906).

As in $\S$\ref{sec:classify}, we define \tauctr\ as the median of \tdep\ within $0.1 \ R_{25}$ and \taudisk\ as the median of \tdep\ between $0.1$ and $0.7 \ R_{25}$. Then, we compare the value of \tauctr\ and \taudisk\ by using a threshold value of 0.26 dex. If log(\tauctr$/$\taudisk) is less than $-0.26$, then that galaxy is in the drop category, and vice versa. For log(\tauctr$/$\taudisk) in between $-0.26$ dex and $0.26$ dex, we assign that galaxy in the flat category.

\begin{figure}
\begin{center}
\includegraphics[width=0.47\textwidth]{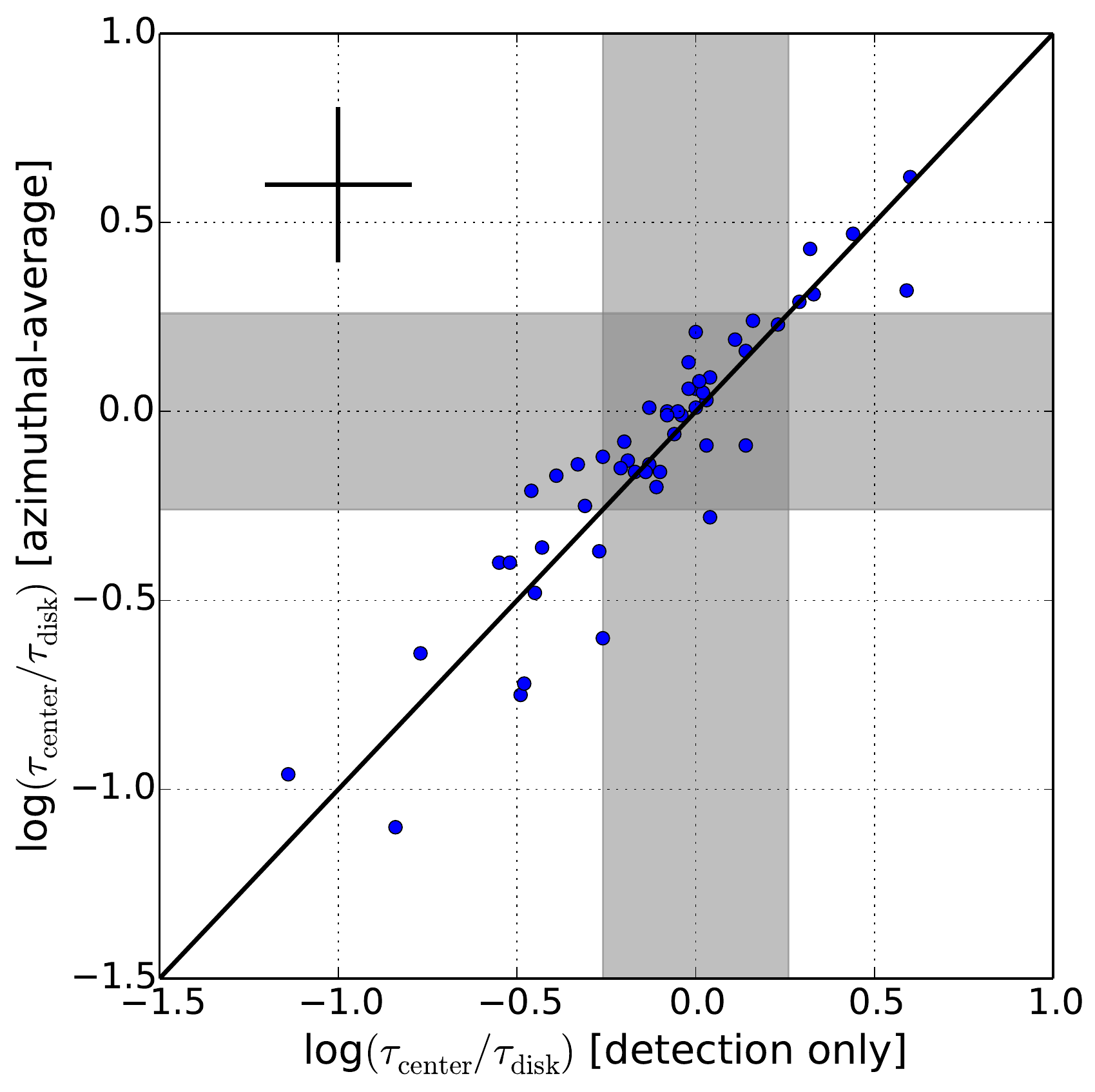}
\end{center}
\caption{A comparison of log(\tauctr$/$\taudisk) between the detection only and after including non-detection as $1 \sigma_{\rm rms}$. The typical uncertainty is about 0.2 dex (the cross sign). The black line is one-to-one correlation. The gray bands mark the spaces of flat category in each method. The number of galaxies in each category is tabulated in Table~\ref{tab:compare_class2}.}
\label{fig:method_comp}
\end{figure}

In Figure~\ref{fig:method_comp}, we plot the values of log(\tauctr$/$\taudisk) that are obtained in $\S$\ref{sec:classify} as the $x-$axis and by including non-detection as the $y-$axis. The relationship between the two values is close to one-to-one relation (black line). This means the inclusion of non-detections almost do not change the results of our analysis in the main text.

Another way to see the effect on non-detections is by comparing the number of galaxies in each group, obtained with the detections only and including the non-detections (as summarizes in Table~\ref{tab:compare_class2}). For detections only, there are 14 galaxies in the drop \tauctr\ group. After the inclusion of non-detections as $1\sigma_{\rm rms}$, only 10 of them stay in the drop \tauctr\ group, while 4 of them are categorized as the flat \tdep\ group. Furthermore, from 32 galaxies in the flat \tdep\ group analyzed using detections only, 30 of them stay in the flat \tdep\ group after the inclusion of non-detections as $1\sigma_{\rm rms}$, while 2 of them are categorized as the drop \tauctr\ group. On the other hand, the number of galaxies in the rising \tauctr\ group is not affected by the inclusion of non-detections as $1\sigma_{\rm rms}$. In total, there are $10+2=12$ galaxies in the drop \tauctr\ group, $30+4=34$ galaxies in the flat \tdep\ group, and 6 galaxies in the rising \tauctr\ group after the inclusion of non-detections as $1\sigma_{\rm rms}$. The numbers of galaxies that stay in the same group are located in the diagonal of Table~\ref{tab:compare_class2}, i.e. $10+30+6=46$ galaxies. If we refer this as "true-positive", then we get a true-positive rate of $46/52 = 88.5\%$, where 52 is the number of galaxies in our sample. For completeness, we also do the same analysis by replacing non-detections with $2\sigma_{\rm rms}$ (Table~\ref{tab:compare_class2}). In this case, the true positive rate reduces to 80.8\%.

\begin{table}
\label{tab:compare_class2}
\centering
\caption{Comparisons of categories between the method in $\S$\ref{sec:classify} (detection only) and azimuthal-average profile by including non-detections (this Appendix).}
\begin{tabular}{| c c | c | c c c c |}
\cline{4-7}
\multicolumn{3}{c |}{} & \multicolumn{4}{c |}{Detection Only} \\
\multicolumn{3}{c |}{} & Drop & Flat & Rise & Total \\
\hline
\multirow{8}{*}{\begin{turn}{90} Include non-detections as \end{turn}} & \multirow{4}{*}{$1 \ \sigma_{\rm rms}$} & Drop & 10 & 2 & 0 & 12 \\
& & Flat  &  4 & 30 & 0 & 34 \\
& & Rise  &  0 &  0 & 6 & 6 \\
& & Total & 14 & 32 & 6 & 52 \\\cline{2-7}
& \multirow{4}{*}{$2 \ \sigma_{\rm rms}$} & Drop & 13 & 7 & 0 & 20 \\
& & Flat  &  1 & 25 & 2 & 28 \\
& & Rise  &  0 &  0 & 4 &  4 \\
& & Total & 14 & 32 & 6 & 52 \\
\hline
\end{tabular}
\end{table}

\section{The Effect of Physical Resolutions} \label{app:res}

The measurement of \tdep\ is known to be scale dependent, that is, the value of \tdep\ changes as a function of physical scale. This difference can be due to the evolutionary effect of star forming regions at scale $\lesssim 0.5$ kpc, where the peaks of CO emission and SFR do not coincidence with each other \citep{kennicutt07,schruba10,kruijssen14}. By using simple models, \citet{calzetti12} found that the scale dependence of \tdep\ is also due to the stochastic sampling of molecular cloud mass functions. However, there is a general trend that \tdep\ reaches an approximately constant value at scales larger than $1-2$ kpc. Interestingly, the central drop of \tdep\ that was reported by \citet{leroy13} occurred at radius $<1$~kpc. Does the central drop of \tdep\ still exists at scales larger than 1 kpc?

To test the scale dependence of \tdep, we degrade the physical resolution of galaxies into 5 scales, from 1 kpc to 3 kpc with an increment of 0.5 kpc. Only galaxies with {\it native} resolutions smaller than a given degraded resolution are included. For example, a galaxy with a native resolution of 0.7 kpc is included in all resolution bins, while a galaxy with a native resolution of 2.2 kpc is only included in degraded resolutions of 2.5 kpc and 3 kpc. In this case, the numbers of galaxies increase from smaller to larger degraded physical resolutions.

\begin{figure}
\begin{center}
\includegraphics[width=0.47\textwidth]{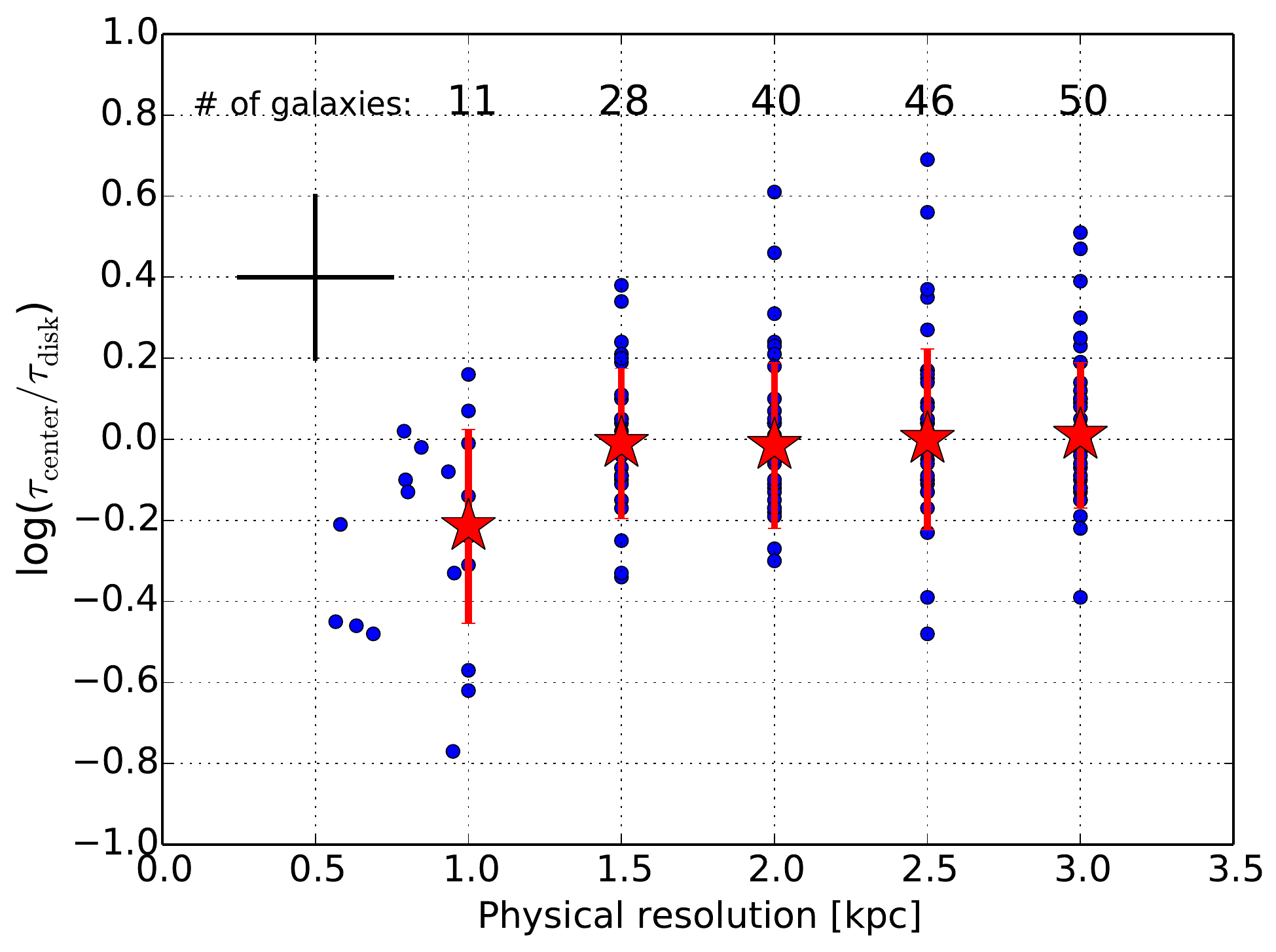}
\end{center}
\caption{The values of \tauctr\ relative to \taudisk\ are plotted against the common physical resolution (blue dots). The red stars are the median values in each physical resolution. The typical uncertainty is marked as a cross. The numbers of galaxies in each resolution bin are indicated on top of the figure. This figure shows that the central drop of \tdep\ is scale dependent and more prominent at scales $\lesssim 1$ kpc.}
\label{fig:test_res}
\end{figure}

The process to make a common physical resolution between galaxies is described below. First, we deproject the EDGE-CALIFA maps of each galaxy by stretching it through its minor axis using an {\tt IDL} function, {\tt GAL\textunderscore FLAT}. During this step, the surface brightness of galaxies are corrected for inclination. Then, we convolve each map to a common physical resolution, corresponding to each degraded resolution, using an {\tt IDL} function, {\tt SMOOTH3D}. Finally, we resample each map using a {\tt MIRIAD} task, {\tt REGRID}, so that each resolution element contains approximately 4 pixels.

In Figure~\ref{fig:test_res}, we show log(\tauctr$/$\taudisk) of each galaxy at various common physical resolution as blue dots. The data points at resolutions smaller than 1 kpc are the values at their native resolution that are included in Figure~\ref{fig:test_res} as comparisons. The red stars mark the median values of log(\tauctr$/$\taudisk) at each resolution. Interestingly, the central drop of \tdep\ is more prominent at resolution $\lesssim 1$ kpc. While there are scatters in the each resolution bin, the median values of log(\tauctr$/$\taudisk) are approximately zero at resolutions larger than 1 kpc. This confirms that the relative values of \tauctr\ with respect to \taudisk\ are indeed scale dependent, and the physical origin of the central drop of \tdep\ is beyond the scale of our data resolution.

If we consider the galaxies with central drop of \tdep\ undergo a nuclear burst of star formation, this implies that the size of that burst is smaller than 1 kpc within the galactic center. A dynamical model of the Milky Way from \citet{krumholz15} predicts that the gravitational instability occurs at scale $\sim 100$ pc in the center. This instability is the result of gas accumulation in the center, driven by the inflow motion due to bar dynamics. Within 17 Myrs time-scale, this gravitational instability leads to a burst of star formation that sweeps out the gas, and then the gas accumulation process restarts again. In this view, our data give a tentative evidence that a burst of star formation may happens in galactic centers.

\end{document}